\documentclass[showpacs,amsmath,amstex,amssymb,mathfonts,pre]{revtex4-1}
\usepackage{graphicx}

\usepackage{tipa}
\usepackage{bm}

\usepackage{amsmath}
\usepackage{amssymb}
\usepackage{amsthm}
\usepackage{amsfonts}
\usepackage{float}
\usepackage{enumitem}
\usepackage{makecell}
\usepackage{multirow}
\usepackage{verbatim}
\usepackage{color}

\begin{document}

\title{Phase Transition in Taxi Dynamics and Impact of Ridesharing}

\author{Bo Yang$^{1,2}$}
\email{yang.bo@ntu.edu.sg}
\author{Shen Ren$^3$, Erika Fille Legara$^6$, Zengxiang Li$^3$, Edward Y.X. Ong$^{2,4}$, Louis Lin$^5$ and Christopher Monterola$^6$}
\affiliation{$^1$ Division of Physics and Applied Physics, Nanyang Technological University, Singapore 637371 .}
\affiliation{$^2$ Complex Systems Group, Institute of High Performance Computing, A*STAR, Singapore, 138632.}
\affiliation{$^3$ Distributed Computing, Institute of High Performance Computing, A*STAR, Singapore, 138632.}
\affiliation{$^4$ School of Applied Engineering and Physics, Cornell University, NY 14850, USA.}
\affiliation{$^5$ Land Transport Division, Ministry of Transport, Singapore, 119903.}
\affiliation{$^6$ School of Innovation, Technology and Entrepreneurship, Asian Institute of Management, Manila, Philippines}

\date{\today}

\date{\today}
\begin{abstract}
We develop a numerical model using both artificial and empirical inputs to analyse taxi dynamics in an urban setting. More specifically, we quantify how the supply and demand for taxi services, the underlying road network, and the public acceptance of taxi ridesharing (TRS) affect the optimal number of taxis for a particular city, as well as commuters' average waiting time and trip time. Results reveal certain universal features of the taxi dynamics with real-time taxi-booking---that there is a well-defined transition between the oversaturated phase when demand exceeds supply, and the undersaturated phase when supply exceeds demand. The boundary between the two phases gives the optimal number of taxis a city should accommodate, given the specific demand, road network and commuter habits. Adding or removing taxis may affect commuter experience very differently in the two phases revealed. In the oversaturated phase the average waiting time is affected exponentially, while in the undersaturated phase it is affected sub-linearly. We analyse various factors that can shift the phase boundary, and show that an increased level of acceptance for TRS universally shifts the phase boundary by reducing the number of taxis needed. We discuss some of the useful insights on the benefits and costs of TRS, especially how under certain situations TRS will not only have economic benefits for commuters, but can also save the overall travel time for the shared parties, by significantly reducing the time commuters spend on waiting for taxis. Simulations also suggest that elementary artificial taxi systems can capture most of the universal features of the taxi dynamics. We give detailed methodologies of the microscopic simulations we employed. The relevance of the assumptions and the overall methodology are also illustrated using comprehensive empirical road network and taxi demand in the city-state of Singapore.
\end{abstract}

\maketitle

\section{Introduction}\label{introduction}

Bridging services such as taxis and shuttles play an important role for balancing the need for comfort, flexibility, and affordability of the transport service in modern cities, enhancing the overall efficiency of urban traffic system\cite{LastMile,EnvironmentTaxi} that also includes the public transportation and private vehicles. In most cities, taxi is the dominant mode of bridging services with the ability to retain most, if not all, of the benefits of owning a private vehicle~\cite{hyang,Wong2001}. Though taxis are more expensive in comparison to public transportation, they are much more flexible and are available on demand. Generally speaking, from the commuters' point of view, whether or not to use taxi services in place of private vehicles or public transportations depends on factors that include cost, the frequency of usage, government policies\cite{Behaviour}, as well as demographic and social-economical variables\cite{ukkusuri}. On the other hand, from a policymaker's perspective, taxis are more desirable than private vehicles because they allow a certain level of sharing that can improve the utility rate of both the vehicle and the transport infrastructure in place\cite{Challenges}. Taxis may also serve as the complementary service to public transportation, especially in places where buses and metros are not readily accessible, and during disruptions or breakdowns of public transportation\cite{LastMile,PublicTrans} . 

Be that as it may, it is also important to note that at the system level, increasing the market share of taxi services can have adverse effects in large and crowded cities. For commuters to choose taxis over public transportation, it generally results in more trips on the road, more energy consumption and green house gas emissions at similar levels as private cars\cite{ShareGreen}. One should particularly note that this is also true when (potential) car-owners choose taxis over private vehicles, especially if taxi ridesharing (TRS) is not an available option, or very few people choose TRS when calling a taxi. If we replace one trip with a personal vehicle from home to office by a taxi ride, it can lead to additional pick-up trips and effectively more vehicles on the road (i.e. taxis that ply on the road without commuters consume road space, and thus contributing to congestions). Hence, reducing vehicle ownership in a city does not always lead to fewer trips on the road, and the contrary may happen when we try meet the demand originally satisfied by private vehicles with taxis without TRS.

It is thus both theoretically interesting and practically useful to have a better understanding of the dynamics of taxis especially with TRS in an urban setting---when the supply and demand of taxi services vary both on spatial and temporal scales\cite{ts2}. Our main methodology in analysing this problem is to use microscopic agent-based modelling and statistical analysis, with a particular focus on the impact of taxi ridesharing (TRS) on taxi dynamics. Some of the main objectives of this paper are to tackle with the following issues: Is there an ideal taxi fleet size for certain city size such that the demand for taxi service is met? How will this number change if commuters are more open to TRS? Do commuters who share rides always sacrifice their total duration of travel (waiting time plus actual trip duration) for a more economic ride? Most of the results we discuss in this paper are quite generic and do not dependent on specific model and algorithm details of a taxi system, as long as the majority of the taxi demand is met with taxi-booking instead of road-side hailing. We focus on taxi booking because with the technological advances it is becoming the mainstream mode for getting taxi services, and also because of its compatibility with TRS. The detailed analysis of the results in this paper lead to better understanding of the benefits and costs of a taxi system, and optimal tuning of such systems from the perspective of both the commuters and the policymakers. 

The organization of the paper is as follows: in Sec.~\ref{simulationa} we describe the main methodologies used for analysing the taxi systems, including a) a large scale agent based simulation platform; b) the edge models we employ when reducing the road network into a graph; c)  construction of algorithms for taxi behaviours; d) approaches for fast query of shortest paths in a complex network. In Sec.~\ref{heuristicmodel} we illustrate an intuitive picture on understanding the interplay between the supply and demand of the taxi services, and how it affects the overall behaviours of the taxi dynamics when different parameters are tuned; in Sec.~\ref{simulatedanalysis} we discuss the extensive numerical results from various artificial and real world road networks, with simulated demand for taxi services, and show interesting statistical results that are universal for different cities and different microscopic details. This includes a well-defined optimal number of taxis given a particular taxi system, as well as differing behaviours in the oversaturated phase (when demand of taxi exceeds supply) from the undersaturated phase (when supply of taxis exceeds demand); in Sec.~\ref{empiricalanalysis}, we focus on the city of Singapore and analyse some crucial behaviours of the taxi system using the empirical demand of taxis throughout a typical day, and draw some interesting statement on how the spatial and temporal variation of the taxi demand, as well as the details of the road network, can affect the commuter travel times and the optimal number of taxis needed; in Sec.~\ref{methodology} we discuss potential applications of the dynamical features of the taxi systems revealed in this paper; in Sec.~\ref{limits} we discuss the limits and generalities of our simulation results, and in Sec.~\ref{conclusions} we summarise our results and discuss about possible future works.

\section{Related work}

For modelling of the taxi system and understanding the interplay of supply and demand of the taxi services, Josep Maria Salanova and Miquel Estrada\cite{WaitingTime} used an agent-based model and empirical taxi demand data to study the taxi dynamics in Barcelona, varying the supply of taxis to determine the socially optimum number of taxis based on the minimum point of the average monetary cost per trip. The waiting time as a function of supply was also considered, and both curves demonstrate a clear decrease in benchmark with increasing supply of taxis when taxi supply is low. Michal Maciejewski et al. investigated how different strategies and distance measures affect various time benchmarks such as the waiting and pick up times of commuters, under different taxi demand to supply ratio in cities such as Mielec\cite{MATSim},Berlin\cite{MATSim2} and Barcelona\cite{MATSim2,MATSim3}. A high and low load regime was defined based on the characteristic of the pick up time against the demand scaling curve, which varied exponentially for the high load regime but linearly for the low load regime. For \cite{MATSim3}, the supply scaling was also considered and a symmetric response to the demand scaling was observed. The focus of the papers was on the effectiveness of different taxi dispatch strategies , while the dynamics behind the transition of the benchmark times against supply and demand were not investigated. The works above did not consider taxi-sharing and the detailed behaviours of commuter waiting and trip time at different supply/demand ratio.

For modelling and optimising the match between taxis and commuters, Ying Shi and Zhaotong Lian\cite{Selfish} modelled the commuter-taxi problem as a double-ended queuing problem and studied the commuters' selfish and selfless threshold to determine the socially optimal allocation of taxis, before discussing possible government intervention methods to achieve and move towards the socially optimal condition. R.C.P. Wong, W.Y. Szeto and S.C. Wong\cite{VacantTaxi} proposed a two-stage modelling approach to predict how empty taxis move when searching for commuters. Stage one estimates the chosen region for taxi drivers to roam about when searching for commuters while stage two looks at the zone chosen by stage one to determine the circulation time and distance of unoccupied taxis in that zone. The model was shown to vastly outperform existing taxi-commuter search models. Fang He and Zuo-Jun Max Shen\cite{EHailing} proposed a spatial equilibrium model that considers the probability of taxi drivers and commuters adopting new e-hailing technology in a well-regulated system and balances the demand and supply of taxi services to estimate how the adoption of e-hailing services will affect the taxi system. Hai Yang et al.\cite{Bilateral} presented an equilibrium model that characterises both the searching and meeting processes between commuters and taxis. A meeting function was defined and used to find the stationary competitive equilibrium and also to explain the differences in decisions by commuters and taxi drivers due to different optimisation aims and the topology of the region considered. Outside of dispatch strategies and taxi dynamics, there has also been increased interests in using GPS data to  predict a city's spatiotemporal taxi supply and/or demand.

For predicting the spatiotemporal distribution of the supply and demand for taxi services, Wang et.al. developed methodologies for stable matching between taxis and commuters with ridesharing\cite{ts3}; Qian et.al looked into the market equilibrium of third-party hailing services and predictions of short-term demand\cite{qian1,qian2}; Moreira-Matias et al.\cite{DemandTaxiStand} constructed a forecasting algorithm to predict commuter demand at taxi stands by combining a non-homogeneous Poisson model with an ARIMA model, showing that inside a 30-minute forecasting timeframe, 78\% of the taxi demand for the city of Porto can be predicted. Benedikt J$\ddot{a}$ger et al.\cite{JagerMunich} analysed the spatiotemporal patterns of taxi data from a local taxi agency in Munich and derived key features that describe the taxi supply and demand characteristics. Following which, a model to predict the demand for complete city districts was suggested and validated with historical data, demonstrating high correlation between predicted and measured data. Xudong Zheng et al.\cite{WaitWhere} looked at the number of unoccupied taxis exiting and the time spent waiting for commuters to board different road segments of a road network to estimate the waiting time at different times of the day on these road segments, from which a recommendation system was developed to direct commuters to locations to wait for taxis to reduce waiting time.

The cases for ridesharing such as increase in travel occupancy\cite{TimeSaving,EnvironmentTaxi}, reduced greenhouse emission\cite{EnvironmentTaxi}, improvement of traffic congestion\cite{Congestion,Congestion2,CongestionUber}, and as an affordable mode of transport to solve the last mile problem\cite{LastMile,Challenges,ts1}, have been well studied in recent literature. Advancements in information technology and their adoptions have also enabled dynamic ridesharing\cite{Challenges,Congestion}, where an automated system matches commuters' travel needs to drivers' availability in an online, real time manner to find some form of optimum\cite{DynamicRS}, providing unprecedented flexibility and convenience. 

In the same vein, much attention has been given to research into factors affecting commuters' decision to share rides\cite{WhyCarpool}\cite{WhyCarpool2}, mechanisms which aid in the adoption of such behaviour\cite{ShareGreen}\cite{TimeSaving}\cite{Incentive}, and methods to quantify the potential for ridesharing \cite{Amey,Shareability1,Shareability2}. Notably, R. Tachet et al. used the concept of a ``share-ability network" to study the potential for ridesharing in the cities of New York, San Francisco, Singapore, and Vienna and demonstrated that remarkably, the share-ability of these cities with diverse sizes and traffic characteristics can all be described by the same scaling law\cite{Shareability2}. Javier Alonso-Mora et al.\cite{Rus} presented a highly scalable anytime optimal algorithm empirically validated using the New York City taxi data. The waiting, travel and total trip time as well as vehicle occupancy rate was studied as a function of different vehicle capacities, vehicle fleet size and maximum waiting and trip delay time, assuming all commuters are open to sharing rides whenever possible, arriving at the conclusion that just 15\% of the current 4-seater taxi fleet in New York City is sufficient to satisfy commuter demand.

\section{Large Scale Microscopic Simulation}\label{simulationa}

Our main methodology for studying the dynamics of the taxi system and the impact of TRS is to use a large scale, microscopic agent based simulation that captures much of the detailed behaviours of both the taxis and the commuters. Such behaviours include the routing of both empty and occupied taxis, boarding and alighting of commuters, booking of taxis, as well as route matching for different commuters sharing part of their rides. The simulation keeps track of the positions and status of all taxis and commuters on a road network at every time step (which can be as fine as one second), and any road network structure can serve as an input to the simulation. The commuters are generated stochastically with a non-trivial spatiotemporal distribution, which can either be synthetic, or based on empirical data. Both the commuter generation and the number of taxis in the simulation can be easily tuned to reflect the interplay between supply and demand.

To analyse the efficiency of the entire taxi system over the road network of potentially the entire city, only statistically robust quantities are used to benchmark the efficiency. These include the average waiting time, average trip time (time spent inside a taxi for the commuters) and average total travel time of the commuters generated over the entire simulation. Given the intrinsic stochastic generation of commuters, choices of commuters and initial status of the taxis (as are the cases in the real world), these averaged quantities are obtained over a large ensemble of simulations of the taxi system for each set of macroscopic features (e.g. the total number of taxis and the spatiotemporal generation patterns of the commuters) and routing algorithms. Detailed explanations will be given for the booking, routing and route-matching algorithms used in our simulations, and how we can optimise these algorithms to not only improve the system efficiency, but also for fast implementations in numerical computations. The latter is also very important for the actual implementations of these algorithms in the real world, given that we are focusing on taxi-ridesharing that is real-time and dynamical. 

\subsection{The Simulation Platform}\label{simulation}

Our large scale simulation platform consists of (1) an underlying road network, (2) the commuters with their origins and destinations denoted by the nodes in the road network, and (3) the taxis that move along the edges of the network from one node to another. A summary of various components of the simulation is listed in Table. I at the end of the paper. We will now introduce each component with more details with a formal set of notations.

\subsubsection{The road network}\label{rn}

Networks are mathematical abstractions of many complex systems, including transportation systems. It is a standard procedure to convert a generic road network into a graph-theoretical representation~\cite{Wong2001, Yang1998, Chen2002, Valenzuela2014}, in which the nodes are intersections or ends of the roads, and the edges are directed. Thus for most of the purposes, a road network corresponds to a graph given by $G\left(\mathcal N,\mathcal A\right)$, where $\mathcal N$ is the set of all nodes in the road network, and $\mathcal A$ is the adjacency matrix that is not necessarily symmetric, with $\mathcal A_{ij}=1$ if there is a physical road connecting $i^{\text{th}}$ and $j^{\text{th}}$ directly, and a vehicle can travel from $i$ to $j$ along that road; otherwise $\mathcal A_{ij}=0$. The adjacency matrix contains information about the topology of the road network, in terms of the accessibility from one location to another.

\begin{figure}[htb]
\centering
\includegraphics[width=16cm]{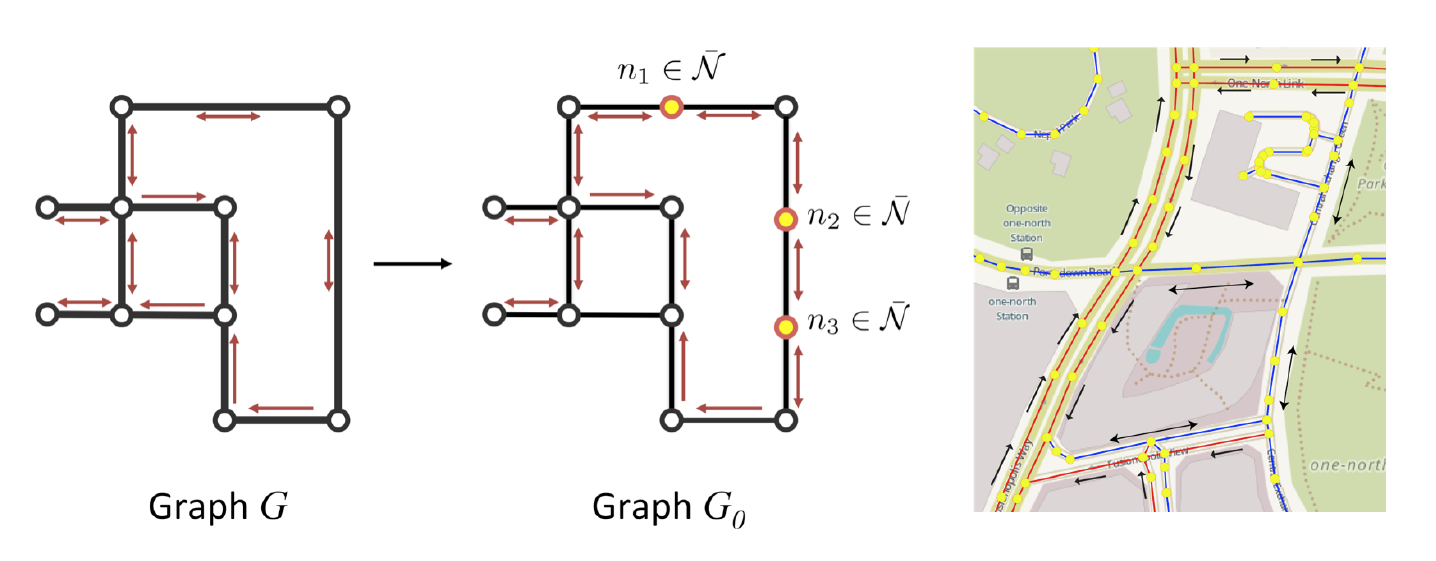}
\caption{In the diagram on the left, we show how $G_0$ is constructed from $G$ (original nodes in $\mathcal N$ are unfilled) by adding nodes ($n_1$, $n_2$, $n_3$) to the original network $G$ such that the resulting adjacent nodes in $\bar{\mathcal N}$ are within $50\sim100$ meters apart. Note that $G_0$ is also weighted, though the edge weights are not explicitly labeled. All the unfilled nodes in $G_0$ are also in $\mathcal N$. The snapshot on the right shows an example of the nodes and directed edges generated from the map of Singapore. Both unidirectional and bi-directional edges can be present, as indicated by the arrows in the figure.}
\label{nodeaddition}
\end{figure}

For the purpose of studying the dynamics of taxis or ride-sharing, our approach is to generate a more fine-grained graph $G_0\left(\bar{\mathcal N},\bar{\mathcal A}\right)$ from $G\left(\mathcal N,\mathcal A\right)$. First of all, $\mathcal N\in\bar{\mathcal N}$, and $\bar{\mathcal N}$ is generated by inserting regularly spaced nodes to each of the edges (except for edges corresponding to highways where boarding or alighting is not possible), so that any two nodes are $50\sim100$ meters apart (Fig.~\ref{nodeaddition}). The additional nodes are inserted so that locations of boarding and alighting can be matched to nearby nodes accurately. The matrix $\bar{\mathcal A}$ is generated from $\mathcal A$ and $\bar{\mathcal N}$, where $\bar{\mathcal A}_{ij}$ is non-zero if for nodes $i,j\in\bar{\mathcal N}$, there is a direct road and the vehicle can drive from $i$ to $j$. $\bar{\mathcal A}$ is also weighted, in which the matrix elements can be equal to the physical distance or average travel time from $i$ to $j$, depending on the purpose of analysis. In this paper, unless mentioned otherwise, the matrix elements are given by the average travel time between two neighbouring nodes. Obviously $\bar{\mathcal A}_{ii}=0$, and if there is no physical roads directly connecting $i$ and $j$, $\bar{\mathcal A}_{ij}=\infty$.

The definition of $G_0\left(\bar{\mathcal N},\bar{\mathcal A}\right)$ allows us to use $G_0$ as  input to our simulation model, where potential commuters book and board taxis from any node in $\bar{\mathcal N}$. The time it takes for taxis to move from one node to another, as well as the routing from two arbitrary nodes, are computed from $\bar{\mathcal A}$. In this way, the simulation does not need any additional geographical information about the city road network, and potential commuters can book taxi almost anywhere along the road, by matching the actual GPS locations of the commuter to the nearest node. Furthermore, the compact structure of $G_0$ allows for more efficient numerical simulation of the taxi dynamics.

For our simulations, the matrix elements of $\bar{\mathcal A}$, or the weighted edges of the graph, are given as the travel time from one node to the other. This allows us to define a time dependent $\bar{\mathcal A}\left(t\right)$, because while the physical distance between any two neighbouring nodes are fixed, the travel time depends on the traffic condition or the average vehicle velocities, which varies between different time of the day (e.g. peak hours v.s. off-peak hours). While this can be easily implemented numerically with the empirical information (i.e. speed band information of each road section at different times of the day), in this paper we use a time-independent $\bar{\mathcal A}$, or a static road network for our statistical analysis. We also do not distinguish between multi-lane road segments from those with single-lane, and there are at most two edges connecting two neighbouring nodes (pointing in opposite directions).

\subsubsection{The generation of commuters}\label{comgen}

Commuters in the simulation can either be generated from empirical data of the actual demand, or artificially based on a particular statistical distribution. Each taxi request is denoted by an abstract object $P^\alpha_t$ generated at time $t$, where the superscript is the index distinguishing between different travelling parties. $P_t^\alpha$ formally contains a pair of nodes indicating the origin and the destination of the trip. It also contains property tags, specifying information including the number of people in the travelling party, as well as other preferences  (i.e. related to taxi-sharing). We do not specify the property tags here as it depends on how detailed the simulation needs to be run. In the latter sections the only property tag we use is the travelling party's acceptance to TRS (it is TRUE if the party accepts TRS).

For randomly generated commuters, at each time step $t$ there is a certain probability of generating $P_t^\alpha$ with a random set of origin and destination, and a random property tag; such probability is related to the rate of commuter generation and can be easily implemented numerically. While there are no restrictions on the property tag, we do require the random origin and destination to be at least five minutes apart. Otherwise the OD pair is considered too short for a taxi trip.

We also have empirical data on the spatiotemporal distribution of the taxi demand in the city of Singapore, thus $P_t^\alpha$ can be fully or partially generated from the empirical data. If certain parts of $P_t^\alpha$ are missing from the empirical data (especially for the property tag), we complement it with appropriate mathematical models. An example of such mathematical models is the OD pairs for the taxi demand: if they cannot be generated from the empirical data, we generate them with a stochastic model(e.g. with normal or poisson distribution both in spatial and temporal domain).

\subsubsection{The dynamics of the taxis}

The taxi simulation is also based on the available empirical data and, in the place of missing data, generated semi-realistically in a stochastic way. The initial positions of the taxis, for example, are usually not available from the data. Thus in most of our simulations, the taxis are placed randomly throughout the road network at the beginning of the simulation. The taxis can also be initialised either as empty or occupied, and different initial conditions can affect the result of the simulation for the first hour or so in some quantitative way. However, such quantitative effects disappears for longer simulations (and we tested with different initial conditions, with or without a warm-up time), and are insignificant especially for the statistical analysis we performed in this paper.

For the movement of taxis, we denote the algorithm on how to choose the pick-up taxi as $\mathcal P$. The booking and route matching algorithm decide which taxi close to the origin of $P_t^\alpha$ would be the one to pick up the travelling party represented by $P^\alpha_t$. The route chosen by the taxi to go from its current position to the pick-up point can either be fitted with empirical GPS data if available, or via specific routing algorithm $\mathcal R_o$. After arriving at the pick-up point, the taxi will stop for a period of $t_l$, the commuter loading time, before moving towards the destination of $P_t^\alpha$ again via $\mathcal R_o$. Once the taxi arrives at the destination and commuters have disembarked, it is a non-trivial problem to model the behaviour of the taxi when it becomes empty again. The routing of the empty taxis are denoted as $\mathcal R_e$. Most of the analysis in this paper do not require accurate modelling of such behaviours, as we have tested extensively with numerical simulations. Here, we just let the taxi roam randomly around until it is booked again. We discuss about this approach in more details in Sec.~\ref{mapproach} and Sec.~\ref{limits}.

Each $P_t^\alpha$ has a flag for taxi-sharing. If the flag is TRUE, the party is open to the possibility of TRS. Given the two parties $P_t^\alpha$ and $P_t^\beta$ with both taxi-sharing flag as TRUE, a route matching algorithm $\mathcal M$ will determine if the two parties will share part or the entirety of their trips. The routing for additional pick-ups and delivery to additional destinations are again determined by $\mathcal R_o$, which can be either empirical or based on specific algorithms. 

\begin{figure}[htb]
\centering
\includegraphics[width=12cm]{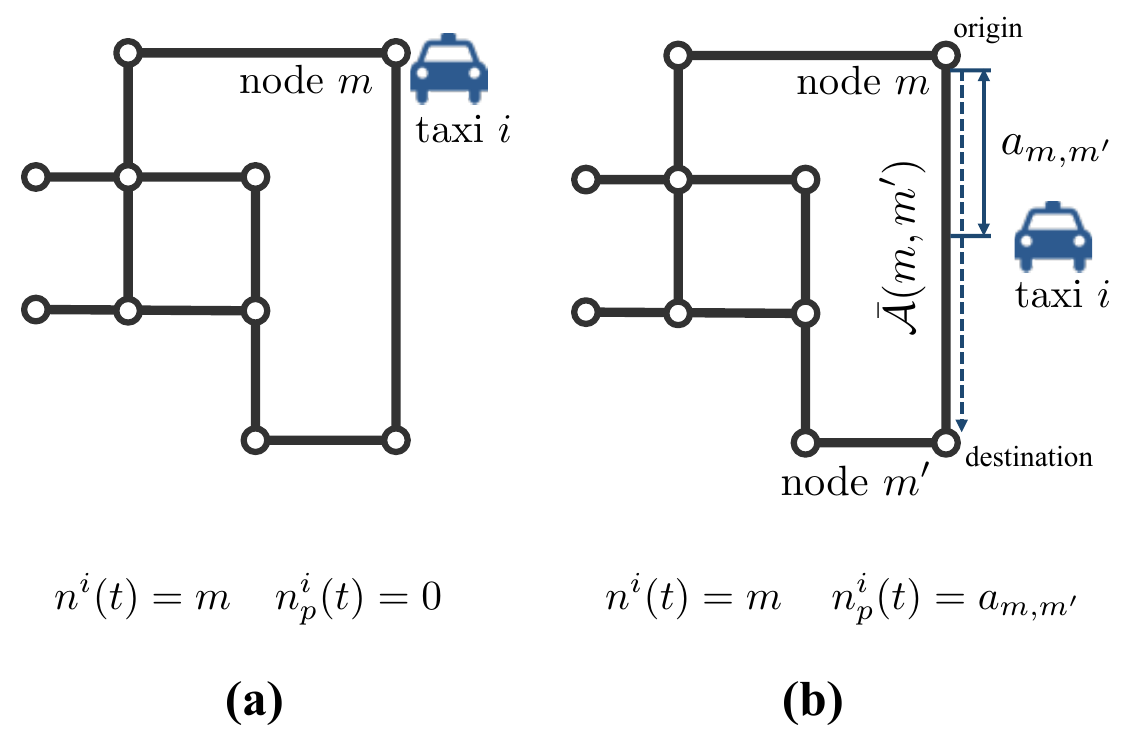}
\caption{Specification of the position of the taxi in the road network. Note that we update the states of all the agents in our simulation at every time stamp, and all the weights of the edges are given as an integer number indicating the number of the time stamps for a vehicle to travel from one node to another.}
\label{taxi-diag}
\end{figure}

For numerical implementation and calculation, the $i^{\text{th}}$ taxi is specified by $n^i\left(t\right)$ and $n_p^i\left(t\right)$. The former gives the node immediately passed by the taxi, while the latter gives the distance of the current taxi's position from that node. Together they specify the position of the taxi in the following way: if the taxi is right at node $m$, then $n^i\left(t\right)=m$ and $n_p^i\left(t\right)=0$; if the taxi is moving from node $m$ to some other node $m'$ along the edge with the weight $\bar{\mathcal A}\left(m,m'\right)$, and at a time distance $a_{m,m'}\le\bar{\mathcal A}\left(m,m'\right)$ away from $m$, then $n^i\left(t\right)=m$ and $n_p^i\left(t\right)=a_{m,m'}$ (see Fig.(\ref{taxi-diag})). Clearly $n^i\left(t\right)=m,n_p^i\left(t\right)=\bar{\mathcal A}\left(m,m'\right)$ and $n^i\left(t\right)=m',n_p^i\left(t\right)=0$ are equivalent. The graph representing the road network is directed, so that there could be two edges connecting a pair of neighbouring nodes $\left(i,j\right)$, with one edge connecting from $i$ to $j$, and the other from $j$ to $i$. Since $n^i\left(t\right)$ gives the node immediately passed by the taxi, the position of the taxi is uniquely determined by $n^i\left(t\right)$ and $n_p^i\left(t\right)$. We also merge different lanes along the same direction into a single edge. Since commuters can only board or alight at nodes (and not in the middle of an edge), the merge does not affect the dynamics of the taxis.

For other properties of the taxis, we can introduce $s^i\left(t\right)$ to denote the status of the taxi, with $s^i\left(t\right)=0$ denoting empty taxis, $s^i\left(t\right)=1$ denoting booked taxis, $s^i\left(t\right)=2$ denoting occupied taxis that are available for taxi-sharing, and $s^i\left(t\right)=3$ denoting occupied taxis not available for taxi-sharing. We also use $s_p^i\left(t\right)$ to denote the number of commuters in the $i^{\text{th}}$ taxi. Other information, such as $P^\alpha_t$ for each travelling party in the taxi, is also stored and updated at each time step if necessary. Once an empty taxi receives a booking, it is assigned a trip to go from its current position to the origin of the commuter, and then from the origin to the destination of the commuter. The actual path is based on the chosen routing algorithm $\mathcal R_o$. Such trip information is stored in $\vec t^i$, which is a vector consisting of an array of nodes arranged sequentially along the path. The vector $\vec t^{ i}$ is only updated when $s^i\left(t\right)$ is changed from $0$ or $2$ to $1$, which is when a booking request is received.

We would like to emphasize here that while there are two major modes of commuter pickups: real-time booking and road-side hailing of taxis, in this paper we focus exclusively on taxis picking up commuters via booking unless otherwise stated. This is because TRS can only be realistically implemented with real-time booking (as we are looking at real-time dynamical taxi-sharing that is not pre-arranged), and it is a major component of analysis in this paper. For the common practice of road-side hailing of taxis, it can be implemented with a non-trivial $\mathcal R_e$ for empty taxis. When the taxi demand is low, taxi-booking is more efficient for available taxis to reach waiting commuters, as compared to road-side pickups (with pseudo-random movement of empty taxis); with high taxi demand, the reverse is true, and we discuss this in more details in Sec.~\ref{limits}. The dynamics of the taxi system with taxi-booking is also quite different from that with road-side hailing, as we will see in Sec.~\ref{simulatedanalysis}. More realistic simulations with a mixture of taxi-booking and road-side hailing will be studied elsewhere.

\subsubsection{A minimalist simulation approach}\label{mapproach}

While our simulation platform is capable of simulating complex taxi and commuter behaviours as well as different routing/pickup/sharing algorithm, here we pursue a minimalist approach and employ the simplest possible simulation algorithms that still retain the essence of taxi dynamics. We are not aiming to quantitatively reproduce the features of the taxis in real cities. This requires a lot of empirical information (e.g. traffic conditions, actual routing of the taxis, detailed spatiotemporal patterns of demand and commuter loading time, as well as the time dependence of the actual number of taxis running on the streets, etc.), even though given such information they can be easily implemented in our codes for more quantitatively accurate predictions. Instead, we aim to capture some of the universal trends of taxi dynamics in a qualitative way and shed light on the intricate interplay between supply, demand and road networks, from the commuters', taxis drivers' and policy-makers' perspective. Following the principle of Occam's razor, it is also useful to show that such universal trends can be captured by the minimalist model (with artificial road networks), as we compare numerical results in Sec.~\ref{simulatedanalysis} and Sec.~\ref{empiricalanalysis}.

Unless otherwise stated, in our simulation each travelling party consists of only one commuter, and a taxi can take two travelling parties at most. Each taxi always moves towards its next destination via the shortest path (implying shortest time, since the edge weights are measured as the time cost) as given by the Dijkstra's algorithm (more details in Sec.~\ref{routing}), so that $\mathcal R_o$ always finds the shortest path between the original node and the final node. Once the taxi reaches its final destination (after dropping off all the commuters in the taxi), instead of modelling its behaviour we make it either roam randomly or remain at the final node until it is booked by another commuter. We can also make the empty taxis move towards some predefined ``hotspots". We have done extensive (though not necessarily thorough) simulations, and it seems different models only affect the results in the following sections in minor and quantitative ways. The universal features of taxis and TRS discussed in this paper are not sensitive to this part of the taxi dynamics. Thus $\vec t^i$ is empty if $s^i\left(t\right)=0$, and $\mathcal R_e$ corresponds to a random walk. The algorithm we use for assigning taxis to commuters is also simple: when a commuter calls a taxi from a certain node, we search through all the positions of available taxis at that time, and assign the taxi closest to the commuter (based on $\mathcal R_o$, or the shortest path) for the pick-up. 

Many route matching algorithms for TRS have been proposed in the literature\cite{TimeSaving,EnvironmentTaxi,EnvironmentTaxi,Congestion,Congestion2,CongestionUber,LastMile,Challenges,DynamicRS}. In this paper, we do not focus on formulating and benchmarking an optimal (in terms of efficiency and ease of implementation) route matching algorithm. The detailed pick-up algorithm ($\mathcal P$) and route matching algorithm ($\mathcal M$) used in our simulations will be constructed in Sec.~\ref{bookingrouting}. Both $\mathcal P$ and $\mathcal M$ can be easily upgraded in our simulations. For actual implementations, it is in general the compromise between efficiency, hardware resources and economics of the scheme. We would like to emphasise here that the qualitative features regarding different phases of the taxi dynamics discussed in this paper is universal; different $\mathcal P$ and $\mathcal M$ only alter certain quantitative features. The latter will not be accurately captured in simulations anyway, as it requires a lot of detailed empirical inputs (e.g. passenger and taxi behaviours, detailed traffic conditions etc.) which is not within the scope of this paper.
\begin{table}[htb]
\begin{tabular}{| c | c | c | c |}
\Xhline{3\arrayrulewidth}
&&&\\
Taxi system components & Notations &  Implementations & Comments\\
& & &\\
\hline
& & &\\
 &  &  & $\bar{\mathcal N}$ is the number of nodes \\
& &Static network, the weights of & at which commuters can be picked up;\\
Road network&$G_0\left(\bar{\mathcal N},\bar{\mathcal A}\right)$&edges are measured in time,& $\bar{\mathcal A}$ can be either static or dynamic,\\
&&assuming constant vehicle velocity of& with edges depending on the traffic conditions\\
&&$60km/h$.& at different time of the day.\\
&&&\\
\hline
&&&\\
Taxi number & $N_T$& Constant for each simulation& $N_T$ can also be time-dependent with \\
&&& more empirical input for realistic simulations.\\
&&&\\
\hline
&&&\\
 &&$P_t^\alpha$ contains only one commuter; & Each object $P_t^\alpha$ is generated at time t,\\
&&the commuter's choice on taxi-sharing& and $\alpha$ is the index of the object;\\
&& is probabilistic, no sharing & each object contains information including\\
Commuter generation&$P_t^\alpha$&preference is specified;& the number of commuters, origin and destinations,\\
&&detour cut-off time same & and if the party agrees to taxi-sharing; it can also\\
&&for all commuters; origin and& contain other properties including preference for\\
&&destinations either randomly picked& taxi-sharing, such as max no. of people in the\\
&&or based on empirical data.& taxi, gender preference, detour time cut-off, etc.\\
&&&\\
\hline
&&&\\
&&  &Every time a $P_t^\alpha$ is generated, $0\le p_s\le 1$ is\\
&&& the probability that the taxi-sharing flag of $P_t^\alpha$\\
Willingness for taxi-sharing&$p_s$&Location and time-independent& is true; it can be time and/or location dependent \\
&&&\\
&&&\\
\hline
&&&\\
Detour cut-off time&$t_c$& $t_c=180\text{sec}$& Can be different for different $P_t^\alpha$\\
&&&\\
\hline
&&&\\
&&&More sophisticated pick-up algorithm\cite{RoutingPickup,RoutingGPS,MeetingPoints}\\
&&&can be implemented; road-side hailing can be\\
Pick-up algorithm&$\mathcal P$& Specified in Sec.~\ref{bookingrouting}&implemented with specified routing algorithm\\
&&&for empty taxis, when it passes nodes \\
&&&with waiting commuters.\\
&&&\\
\hline
&&&\\
commuter loading time&$t_l$&$t_l=0$&$t_l$ can be position, time and/or\\
&&& party-size dependent\\
&&&\\
\hline
&&&\\
Route matching algorithm&$\mathcal M$&Specified in Sec.~\ref{bookingrouting}&Different algorithms can be implemented\cite{RoutingTimeDep,RoutingDynamic}\\
&&&\\
\hline
&&&\\
Occupied or booked taxi&$\mathcal R_o$&Dijkstra's algorithm&Different algorithms or empirical input.\cite{RoutingDynamic,RoutingOccupied}\\
routing algorithm&&&\\
&&&\\
\hline
&&&\\
Empty taxi routing&$\mathcal R_e$&Random roaming for empty taxis&Different algorithms or empirical input.\cite{RoutingEmpty}\\
algorithm&&&\\
\hline
\end{tabular}
\caption{Components, notations and both of their possible and actual implementations of the taxi system.}
\label{t}
\end{table}

\subsection{Edge models for graph representations of road network}

The road network is one of the most important components of the transportation system. Methods of modelling the networks are well-studied, and most of the approaches involve the abstraction of the road network into a simple graph. In these approaches, it is usually assumed that there exists a best route between any two nodes in this graph. A recent interesting work\cite{hamza} shows that for many realistic problems, every road segment needs to be assigned with several attributes. This implies that the best route between any two nodes could be an ill-defined concept, either because it is not unique due to various different trade-offs, or because it is not useful for practical applications. A multigraph approach was proposed, which is particularly useful for the Vehicle Routing Problem with Time Windows (VRPTW). Though the nature of the networks studied in Ref.\cite{hamza} is not exactly the same as the road networks for taxi systems, the latter is also a multi-faceted system requiring more than a simple graph approach in which edges weights only correspond to physical distances.

\begin{figure}[htb]
\centering
\includegraphics[width=11cm]{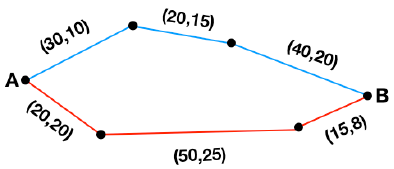}
\caption{Two potential routes from Node A (origin) to Node B (destination), where for each edge we have $\left(f_1,f_2\right)$: $f_1$ gives the time cost, and $f_2$ gives the fuel cost. If we only consider the time cost, the bottom route will be chosen; if we only consider the fuel cost, the top route will be chosen. If we define a utility function $u=f_1+\lambda f_2$ for the edges, then the top route will be chosen for $\lambda>0.625$, while the bottom route will be chosen for $\lambda<0.625$.}
\label{fig18}
\end{figure}

Given that there is only one type of moving agents (the taxis, with a fixed total number) roaming in the road network, and only one type of nodes (all nodes are equivalent, modelling locations where commuters can board or alight taxis), we still use a graph representation in this paper, which should be sufficient especially since the adjacent nodes are $50\sim 100$ meters apart (see Sec.~\ref{rn}). On the other hand some elements of the multi-graph approach can be incorporated to make the formulation as general and flexible as possible for various different applications. A number of factors play important roles in taxi systems, including the travel and waiting time, fuel cost, road tolls, driver earnings and commuter satisfactions, etc. All of them are directly affected by the routing of taxis, and can be captured by the proper modelling of the edge weights, with the following most general representation:
\begin{eqnarray}\label{multi}
\mathcal F:E\rightarrow \mathcal R^n,\qquad f_k\left(\left\{i,j\right\}\right)=r_k,\quad k=1,2,\cdots n
\end{eqnarray}
where $E$ is the collection of all edges of the road network indexed by $\left\{i,j\right\}$, the \emph{ordered} pair of neighbouring nodes, and $\mathcal R^n$ is the $n$-dimensional real space, with $n$ being the number of factors considered for the taxi system. As a simple example, $f_1$ maps the edges to the average travel time between the neighbouring nodes, and $f_2$ maps to the average fuel cost between the two neighbouring nodes (see Fig.(\ref{fig18})).

For actual implementations of vehicle routing, the first option is to have $n$ sets of routing between two nodes, each set corresponding to the shortest path calculated from the edge weights given by $f_k\left(\left\{i,j\right\}\right)=r_k$ for a particular $k$. One should also note that in practice different factors given by different $f_k$ are highly correlated with each other. If the travel time of a particular route between the two nodes is large, in most cases the fuel cost for that route will be large as well. Thus we expect that using different $f_k$ can lead to the same routing between many different pairs of nodes. All such routings can be recommended to the drivers for them to choose from, contingent upon other real-time and human factors.

Another option is to define an overall utility function over the directed pair of neighbouring nodes, mapping $\vec f=\left(f_1,f_2\cdots, f_k\right)$ to a real number by $\mathcal U\left(\vec f\left(\left\{i,j\right\}\right)\right)=u\left(\left\{i,j\right\}\right)$. For example if $n=2$, we can have
\begin{eqnarray}\label{utility}
u\left(\left\{i,j\right\}\right)=f_1\left(\left\{i,j\right\}\right)+\lambda f_2\left(\left\{i,j\right\}\right)
\end{eqnarray}
where $\lambda$ is the tuning parameter measuring the relative importance of $f_1$ and $f_2$. In this way, even though the routing between any two nodes in the network is unique, all different attributes of the edges (or the road segments) are involved, and the chosen route can change when a different $\lambda$ is used, as we illustrate in Fig.(\ref{fig18}). 

We also would like to emphasise that the general analysis of the qualitative behaviours of the taxi dynamics in this paper does not depend on different ways we model the edges in the graph. Different approaches lead to different values assigned to each of the edges, but our results are universal for any (potentially randomly assigned) edge values, encoded in the adjacency matrices $\mathcal A_{ij}$ we input into our simulation platform. It also shows that our simulation platform can be easily adapted to a wide family of system optimisations (by specifying Eq.(\ref{multi}) and Eq.(\ref{utility}), via the abstraction we use in the simulation (see Table.~\ref{t}). In particular for all the simulations we performed in this paper, the edge model is given by a simple utility function that puts an overwhelming weight on the travel time between the two neighbouring nodes, so it basically just measures the average travel time between the nodes. This is useful especially when comparing to the model with physical distances as edge weights, because with travel time the edge weights can fully account for speed limits, traffic conditions and even traffic light patterns. It is also the most relevant choice given our focus on analysing the impact on commuter waiting time. The generalisation to other related scenarios with Eq.(\ref{multi}) and Eq.(\ref{utility}), as well as a multigraph approach as discussed in\cite{hamza}, is currently work in progress and will be presented elsewhere. 

\subsection{Booking and Route-matching Algorithms}\label{bookingrouting}

For taxi-booking with the possibility of ride-sharing, the pick-up algorithm $\mathcal P$ decides which empty or occupied taxi is assigned to pick up a waiting commuter, and the route matching algorithm $\mathcal M$ decides which occupied taxi can be an eligible candidate for pick-up. Thus the route matching algorithm decides which subset of occupied taxis can \emph{potentially} serve a particular waiting commuter. On the other hand, the pick-up algorithm chooses the best taxi, from this subset of occupied taxis \emph{and} all the empty taxis, for the actual pick-up and service of this commuter. We first describe the details of $\mathcal M$ implemented in our numerical simulations. 

{\it Route Matching Algorithm--} When a commuter is generated with $P_t^\alpha$ (see Sec.~\ref{comgen}), there is a certain predefined probability $p_s$ that the taxi-sharing flag of $P_t^\alpha$ is TRUE, indicating the commuter is open to taxi-sharing. We now focus on the case where the taxi-sharing flag is TRUE, with the origin of the corresponding $P_t^\alpha$, which is waiting for the taxi, given by the node $O_\alpha$ and destination $D_\alpha$. Let $\mathcal T_0\left(t\right)$ be the collection of taxis at $t$ with $s^i\left(t\right)=0,i\in\mathcal T_0\left(t\right)$, and $\mathcal T_s\left(t\right)$ be the collection of taxis with $s^i\left(t\right)=2,i\in\mathcal T_s\left(t\right)$. For each taxi $i\in\mathcal T_s\left(t\right)$, it is moving from its current position $n=n^i\left(t\right)$ to the destination of the commuter in the taxi, which we denote as $D_\beta$. Let $t_{AB}$ be the time it takes for the taxi to go from node $A$ to node $B$ based on the shortest path as measured in travel time (i.e. implemented with $\mathcal R_o$), and we define $t_{ABC}=t_{AB}+t_{BC}, t_{ABCD}=t_{AB}+t_{BC}+t_{CD}$. We specify the following conditions:
\begin{eqnarray}
&&t_{nO_\alpha D_\beta}-t_{nD_\beta}<t_c\label{c1}\\
&&t_{nO_\alpha D_\alpha D_\beta}-t_{nD_\beta}<t_c\label{c2}\\
&&t_{O_\alpha D_\beta D_\alpha}-t_{O_\alpha D_\alpha}<t_c\label{c3}\\
&&t_{O_\alpha D_\beta D_\alpha}>t_{O_\alpha D_\alpha D_\beta}\label{c4}
\end{eqnarray}

\begin{figure}[htb]
\centering
\includegraphics[width=12cm]{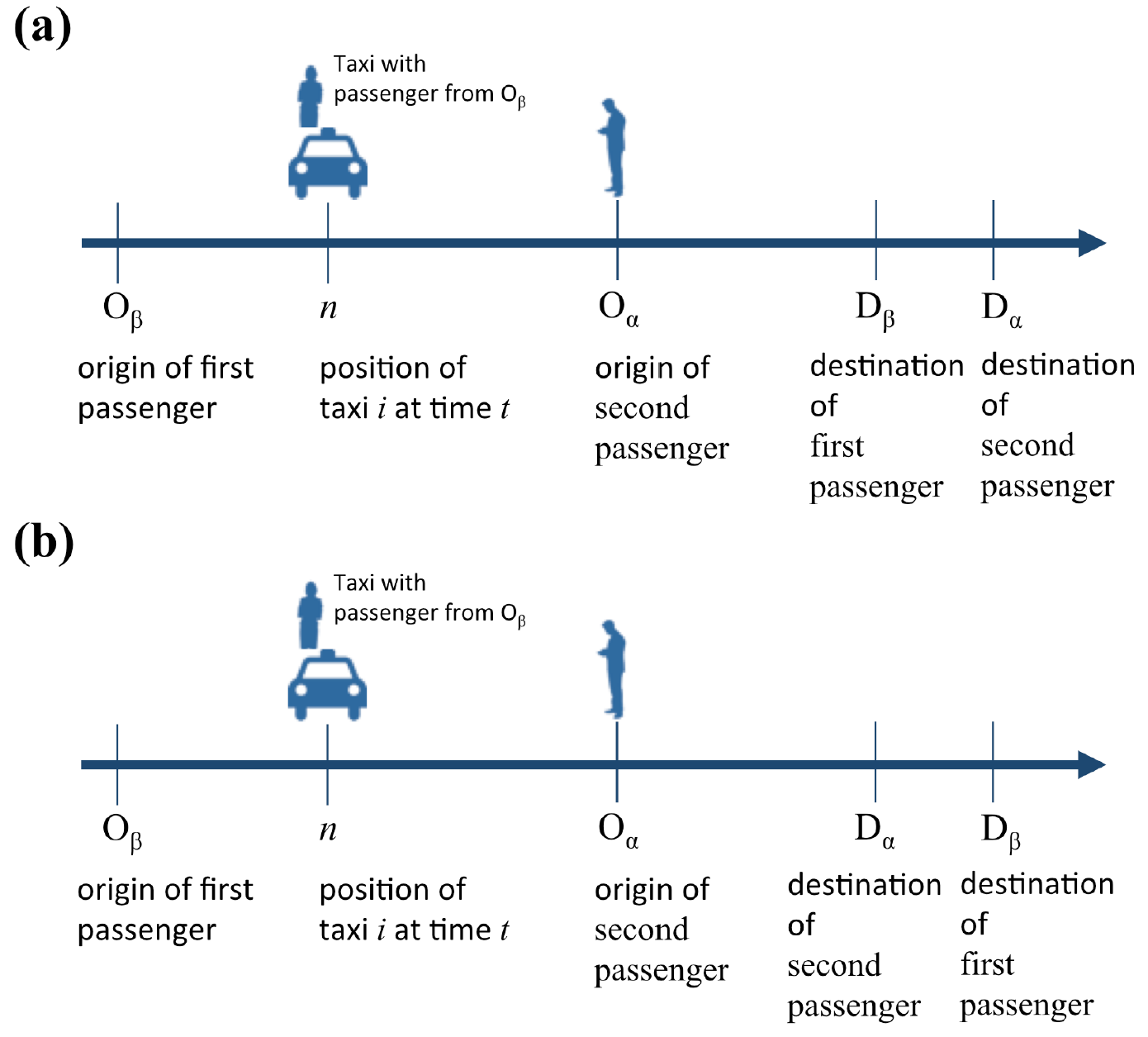}
\caption{Schematic drawing of the order of pickups and dropoffs after route-matching; a).$n\rightarrow O_\alpha\rightarrow D_\beta\rightarrow D_\alpha$; b). $n\rightarrow O_\alpha\rightarrow D_\alpha\rightarrow D_\beta$}
\label{pickup}
\end{figure}

Here $t_c$ is the tuning parameter which indicates the cut-off of the detour time, and we set it as $180$ seconds in the simulations (which is around $15\%$ of the average trip time of $20$ minutes). Both commuters involved in the taxi-sharing should not spend their respective time in the taxi longer by $t_c$, as compared to the time it takes for each of them to go from their respective origin to destination without taxi-sharing. This can be enforced with the route matching algorithm $\mathcal M$ specified as follows (also see Fig.(\ref{pickup})):
\begin{itemize}
\item If Eq.(\ref{c1}),Eq.(\ref{c2}) are satisfied and Eq.(\ref{c3}) is not satisfied, taxi-sharing can occur with the only possible option of the routing $n\rightarrow O_\alpha\rightarrow D_\alpha\rightarrow D_\beta$; 
\item If Eq.(\ref{c1}) is satisfied but Eq.(\ref{c2}) is not, taxi-sharing can only occur if Eq.(\ref{c3}) is satisfied, with the routing of $n\rightarrow O_\alpha\rightarrow D_\beta\rightarrow D_\alpha$; 
\item If all Eq.(\ref{c1}), Eq.(\ref{c2}) and Eq.(\ref{c3}) are satisfied, both routing are possible, and we choose the one that minimises the overall travel time: if Eq.(\ref{c4}) is satisfied, we choose the routing of $n\rightarrow O_\alpha\rightarrow D_\alpha\rightarrow D_\beta$, otherwise we choose the routing of $n\rightarrow O_\alpha\rightarrow D_\beta\rightarrow D_\alpha$.
\end{itemize}

We now define $\mathcal T'_s\left(t\right)\in\mathcal T_s\left(t\right)$, and $\mathcal T'_s\left(t\right)$ contains all taxis satisfying one of the above conditions. The pick-up algorithm $\mathcal P$ will be specified as follows: if the taxi-sharing flag is FALSE, the nearest taxi in $\mathcal T_0\left(t\right)$ will come for the pick-up. If the taxi-sharing flag is TRUE, the nearest taxi in $\mathcal T_0\left(t\right)\bigcup\mathcal T'_s\left(t\right)$ will come for the pick-up. The key feature here is that if the commuter chooses to be open to taxi-sharing, there are potentially more available taxis around for the nearest one to be selected (including empty taxis and occupied taxis satisfying the route-matching algorithm). This could lead to shorter waiting time as compared to commuters who choose not to accept taxi-sharing.

With all the relevant algorithms properly defined, the complete dynamics of individual taxis at time step $t$ can be uniquely specified. Given that $n^i\left(t-1\right)=m$, and if $s^i\left(t\right)>0$ we have $\vec t^i\left(k\right)=m$, or the $k^{\text{th}}$ entry of $\vec t^i$ is $m$, we have:
\begin{enumerate}[label=(\alph*)]
\item If $s^i\left(t\right)=0$, the $i^{\text{th}}$ taxi will move based on our choice of trivial $\mathcal R_e$ (in our case a random walk);
\item If a booking request is sent to the $i^{\text{th}}$ taxi based the pick-up algorithm $\mathcal P$, $s^i\left(t\right)=1$, and $\vec t^i$ is updated accordingly;
\item Movement of taxis: if $s^i\left(t\right)>1$, $n_p^i\left(t\right)=n_p^i\left(t\right)+1$; if $n_p^i\left(t\right)=\bar{\mathcal A}\left(m, \vec t^i\left(k+1\right)\right)$, $n^i\left(t\right)=\vec t^i\left(k+1\right),n_p^i\left(t\right)=0$;
\item Commuter boarding: if $s^i\left(t\right)=1$ and $n^i\left(t\right)$ is the origin of the commuter waiting to be picked up, $s^i\left(t\right)=2$ if the commuter is accepting taxi-sharing, or $s^i\left(t\right)=3$ if the commuter does not accept taxi-sharing. The number of commuters in the taxi is also updated as $s^i_p\left(t\right)=s^i_p\left(t\right)+1$;
\item Commuter alighting: if $s^i\left(t\right)=2$ or $s^i\left(t\right)=3$ and $n^i\left(t\right)$ is the destination of one of the commuters in the taxi, $s^i_p\left(t\right)=s^i_p\left(t\right)-1$. If $s^i_p\left(t\right)=0$, $s^i\left(t\right)=0$, otherwise $s^i\left(t\right)=2$;
\end{enumerate}

For Step (b), the booking request can be sent to taxis with status $s^i\left(t\right)=0$ or $s^i\left(t\right)=2$. In the former case the taxi is empty, and in the latter case the taxi is occupied with one commuter who is open for taxi-sharing. The trip information of the taxi, $\vec t^i$, is only updated at Step (b). Once $s^i\left(t\right)$ is set to $1$, $\vec t^i$ is reset with the shortest path from the taxi's current position to the origin of the commuter who sends the booking request, followed by destination(s) of the commuters in the taxi, based on the route matching algorithm $\mathcal M$ if necessary.

{\it Pick-up Algorithm--} For the pick-up algorithm $\mathcal P$, both empty taxis and occupied taxis that are eligible based on $\mathcal M$ are possible candidates. We use the simple criterion that among all possible candidates, the taxi that is closest to the booking commuter (via the shortest path routing algorithm described in Sec.~\ref{routing}) is chosen for the pick-up. One important tuning parameter for the pick-up algorithm is the maximum range of booking $ R$. Since we define the edges of the road network in units of time for taxis to travel from one end to the other, $ R$ is also measured in units of time: only candidate taxis that can reach the commuter within $R$ can come for the pick-up. For simplicity, in Sec.~\ref{simulatedanalysis} we always set $R\rightarrow\infty$, so that as long as there are possible candidates somewhere in the road network, the closest one will always come for the pick-up no matter how far away it is. In Sec.~\ref{methodology} we will also briefly look at how tuning $R$ can affect the efficiency of the taxi system.

\subsection{Efficient Implementations of Routing Algorithms}\label{routing}

Both for simulation and for actual implementation, the most frequently used algorithm is the shortest path routing algorithm to find the optimal routes that span origins and destinations with minimum costs. One of the most commonly applied methods is to conceptualise road networks as directed weighted graph and to find shortest path (based on the weights of the edges) between two given nodes indicating origin and destination in the graph. Since in our paper the edge weights are defined as the time it takes for the taxis to travel between two neighbouring nodes, the algorithm gives the shortest time it takes for the taxis to move one node to another in the network. However, in most of the urban applications and also in this work, it is rather impractical to apply Dijkstra's algorithm as a greedy method (which incurs computational time at one to two orders of magnitude of seconds for each query) when the number of nodes is large. It is thus of utmost importance to construct efficient shortest path routing algorithms, both for generic networks and for urban road networks. For the latter, we can further boost the efficiency of the algorithm by exploiting certain special features of the network.

Using the Singapore road network conceptualised as weighted directed graph $G_0\left(\mathcal N, E\right)$ as an example, where the weight function $c: E \longrightarrow \mathbb{R}^{+}$ represents the time cost traveled on each edge, the routing algorithm is to find the exact minimal cost path as the optimal route and to return the cost between origin-destination node pair $i$ and $j$ if vertices $i,j\in\bar{\mathcal N}$, denoted as $\mathcal A_{ij}$. We use a customised version of the most efficient shortest path algorithm adopting this surprisingly simple policy called Contraction Hierarchy in our simulations and implementations. The details of the algorithm are illustrated in \cite{CH}.

The most important concept in Contraction Hierarchy is to iteratively rank the importance of the nodes of graph, contract the least important node one at a time and add a short-cut to preserve the shortest path. Theoretically, Contraction Hierarchy can work on any type of node ordering, though different ordering will affect the algorithm efficiency. The optimal ordering of nodes is an NP-hard problem \cite{PreprocessingCH}. For road networks with innate hierarchies (roads are constructed such as expressways, arterial roads, secondary roads and minor roads which have different levels of importance in road network), we develop a heuristic node ordering customised for Singapore road network by taking the road categories into account. The selection of node orders is achieved by a priority queue where the next node to be contracted is determined by a linear combination of priority measures including edge differences, number of contracted neighbours, shortcut cover and node level listed in \cite{CH}, as well as road category which is available for each edge in the Singapore road network. There are six road categories in Singapore - expressway, major arterial road, arterial road, minor arterial road, local assessment and others - in a descending order of priority. The adoption of customised node ordering and contraction hierarchy in our implementation has achieved around tens to hundreds of microseconds for each routing query.

Further boost to the algorithm efficiency can be achieved by pre-calculating the most computationally intensive part of the algorithm, at the cost of large memory storage. For a static road network with adjacency matrix $\mathcal A_{ij}$, we can calculate the shortest distance matrix $\mathcal D_{ij}$, and the shortest path matrix $\vec{\mathcal P}_{ij}$, and store them in advance for the dynamical simulations. Here $\mathcal D_{ij}$ gives the shortest distance (based on the edge weights) between the $i^{\text{th}}$ and $j^{\text{th}}$ node, while $\vec{\mathcal P}_{ij}$ is the vector of the shortest path connecting the $i^{\text{th}}$ and $j^{\text{th}}$ node. Once these two matrices are obtained, subsequent simulations or implementations can just access those matrix elements, instead of recomputing the shortest distance/path repeatedly. Even for dynamical road network in which the edge weights are time dependent, we can pre-calculate $\mathcal D_{ij}$ and $\vec{\mathcal P}_{ij}$ based on average values of the edge weights, so that the real time computations on the fly could be much faster. The speed-up of the routing algorithms is indispensable for the large scale statistical numerical analysis that has been carried out in this paper.  

\section{Heuristic Model for Taxi Services}\label{heuristicmodel}

It is useful to understand the dynamics of taxis, especially their interaction with potential commuters, with a heuristic and intuitive model that captures the essential qualitative features of how such interaction will affect the travel time and waiting time of both commuters and vehicles. What we present here is a conceptual model based on the supply of and demand for taxi service. Such a model can be helpful in interpreting the quantitative and qualitative numerical analysis, which we will detail in the next few sections.

Taking the urban traffic system as a whole, the rate of demand for taxi service, as measured from the number of calls of taxi per unit time, can be denoted by $\mathcal D\left(t\right)$. The integration of $\mathcal D\left(t\right)$ over time gives the total demand. We formulate the supply of the taxi service by the maximum possible rate of commuters that can be picked up by the taxis, $\mathcal S\left(t\right)$, given the number of taxis available. The integration of $\mathcal S\left(t\right)$ over time gives the maximum number of commuters that can be picked up. Note that $\mathcal S\left(t\right)$ describes the capability of taxis to cater to the potential demand for taxi services. The actual rate of pick-up, $\mathcal S^*\left(t\right)$, could be smaller if the actual rate of demand is small, i.e. there are no people waiting on the roads even with available taxis around. 
\begin{figure}[htb]
\centering
\includegraphics[width=0.8\textwidth]{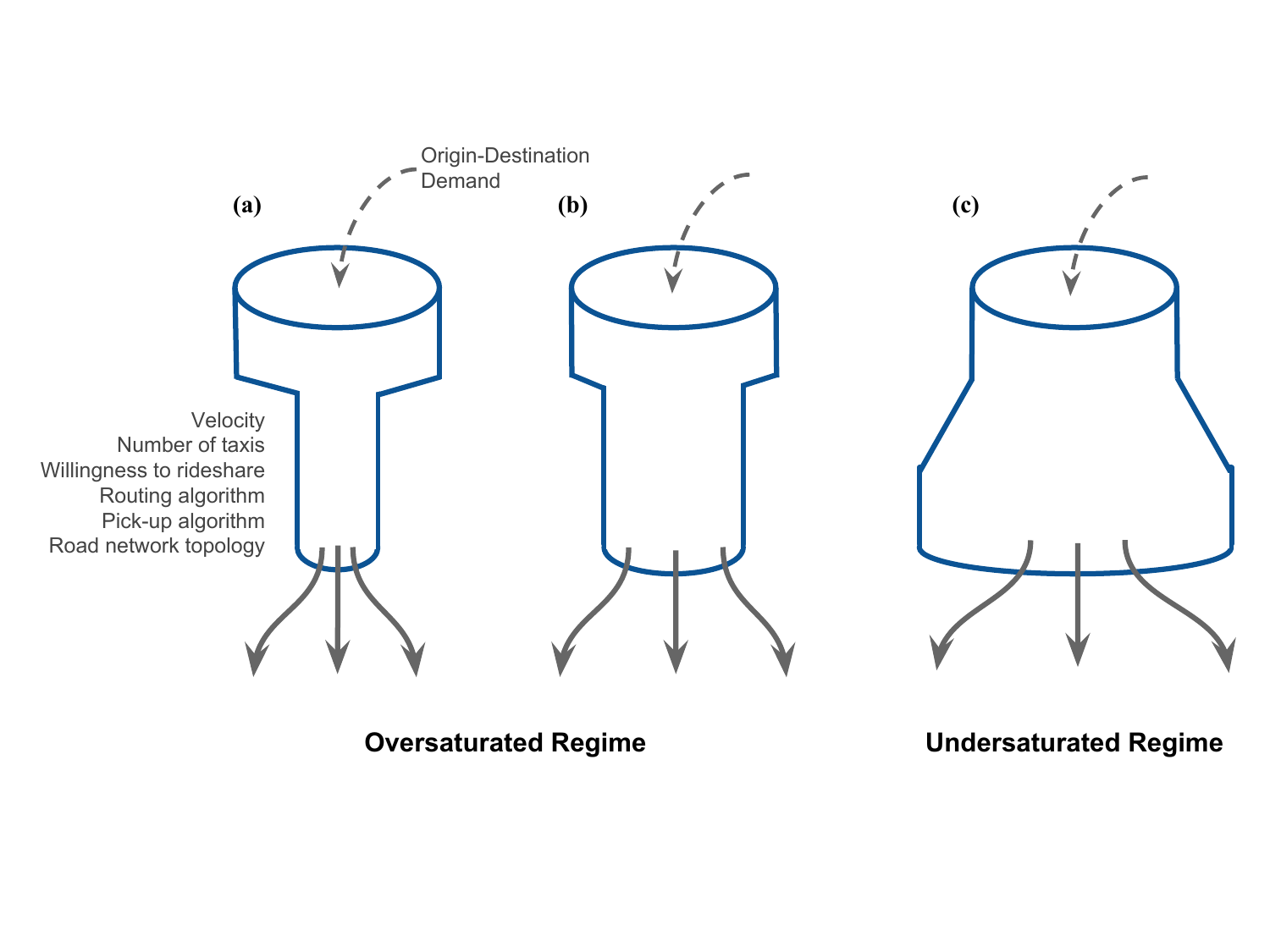}
\caption{The heuristic picture of the interplay between supply and demand of the taxi system. The demand of the taxi services is analogous of the inflow of the ``funnel" representing the taxi dynamics, and is related to the rate of commuter generation $P_t^\alpha$. The outflow from the ``funnel" depends on many factors including $N_T, p_s, \mathcal P, t_l, \mathcal M, \mathcal R_o, \mathcal R_e$ as well as the road network topology given by $G_o\left(\bar{\mathcal N},\bar{\mathcal A}\right)$ itself. In (a) and (b), the system is \textit{over-saturated} (demand is greater than supply), so that the ``water level" in the funnel will keep rising, leading to exponential increase of commuter waiting time; while in (c) the system is \textit{under-saturated}, and there is no accumulation of ``water in the funnel". }
\label{heuristic}
\end{figure}

We can now heuristically divide the dynamics of taxi service into two broad cases. The first case is when there is not enough supply for the demand, formally given by $\mathcal D\left(t\right)\gtrsim\mathcal S\left(t\right)$. In this case, the rate of picking up commuters is lower than the rate of new calls for the taxis. This implies that when a person calls for a taxi, there are no available taxis to come picking him or her up. He or she will have to wait for some occupied taxi to deliver the commuters first, before coming for the pick up. Moreover, the longer that $\mathcal D\left(t\right)\gtrsim\mathcal S\left(t\right)$ persists, the longer the waiting time the new commuters will have to suffer, as one can see from the schematics in Figs. \ref{heuristic}a and \ref{heuristic}b. We should also note that in this case $\mathcal S\left(t\right)\simeq\mathcal S^*\left(t\right)$.

The second case is when there is enough (\emph{see} Fig.~\ref{heuristic}c), or oversupply of the taxi service, formally given by $\mathcal D\left(t\right)\lesssim\mathcal S\left(t\right)$. In this case, $\mathcal S\left(t\right)=\mathcal S^*\left(t\right)$ if there are people waiting for the taxis (this could be because $\mathcal D\left(t\right)\gtrsim\mathcal S\left(t\right)$ has persisted for a while). When most of the waiting people have been picked up, we will have $\mathcal S^*\left(t\right)<\mathcal S\left(t\right)$ and $\mathcal S^*\left(t\right)\simeq \mathcal D\left(t\right)$. 

The demand for taxi, $\mathcal D\left(t\right)$ is in principle empirical, while $\mathcal S\left(t\right)$ characterizes the capacity of the taxi services in a city to clear away waiting commuters. In analogy to the capacity of the road for traffic and the capacity of intersection control (\cite{ResisFactors,WhyCongest,TaxiTax}), $\mathcal S\left(t\right)$ depends on many factors. First and foremost, one would expect $\mathcal S\left(t\right)$ to increase when the number of taxis running on the street increases, assuming the traffic condition is not severely exacerbated by the increase of vehicle numbers/trips on the road. It also depends on the average length of trips for the commuters. If the average travel time between the origin and destination is longer,  taxis are occupied for a longer time, leading to fewer available taxis for pick-ups. Taxi service capacity also depends subtly on the underlying road network and the specific patterns of the OD pairs. As we will also focus in Sec.~\ref{simulatedanalysis}, the introduction of TRS can also strongly increase  taxi service capacity, given the same number of taxis on the road.

\section{Numerical Simulations and Analysis}\label{simulatedanalysis}

While many different types of simulations can be performed for a road network with taxis and commuters, we focus on both qualitative and quantitative analysis of the behaviours of the taxis and their impact on the commuters, based on our model and using the framework introduced in Sec.~\ref{heuristicmodel} as a general guide. From the policymakers' perspective, given a specific demand for taxi services, it would also be useful to study ways to reduce the number of taxis, and more importantly reduce the number trips on the road, without affecting the quality of service (i.e. making commuters wait longer, etc.). In this section, we will first get some intuitions of these questions with simulations on an artificial network, followed by a more quantitative assessment using the road network of the city of Singapore.

\subsection{Square Lattice Network}

\begin{figure}[htb]
\centering
\includegraphics[width=.5\textwidth]{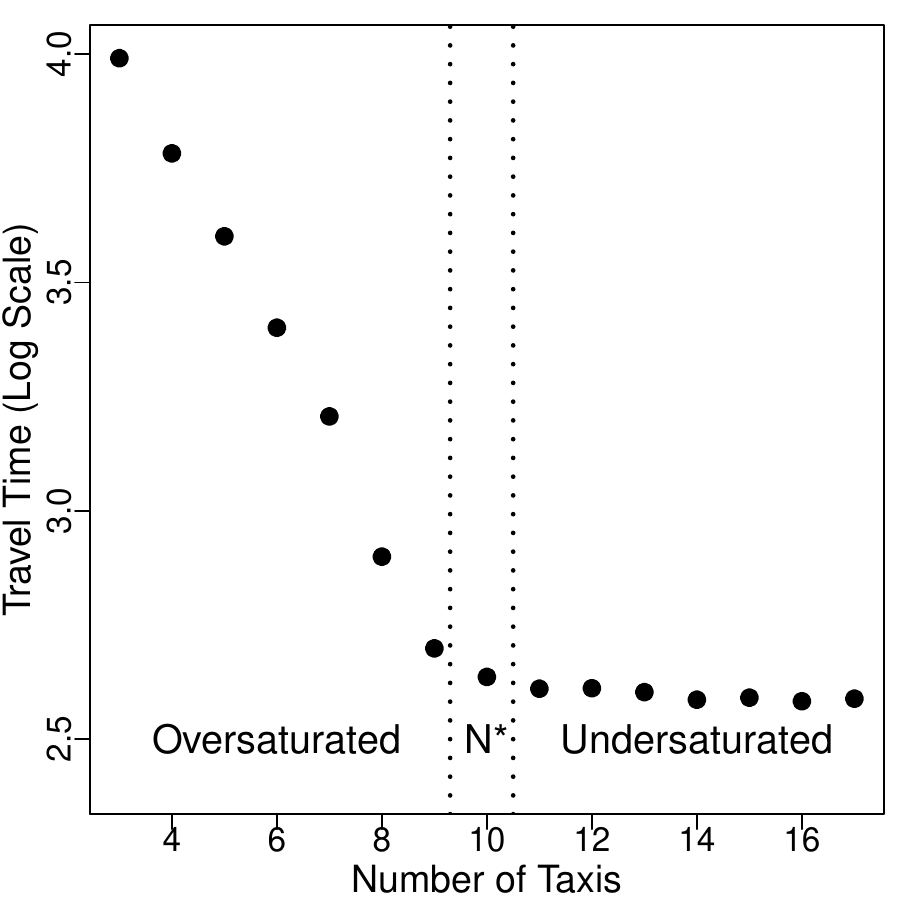}
\caption{A sample plot of the log of the average total travel time (the sum of the average waiting time and the average trip time) as the function of the number of taxis, given a fixed demand of one taxi request every one minute. Each data point corresponds to a total simulation time of $10000$ sec.}
\label{f1}
\end{figure}

It is instructive to first study the artificial road network with simulated transportation supply and demand, to understand some of the general features of the taxi dynamics, especially when TRS is enabled. This is particularly useful if we can find features that are universal, independent of the details of the road network and the patterns of the transportation demand. As a first example, we start with a regular road network in the form of a square lattice with a total of $10000$ nodes. The edges between neighbouring nodes are bi-directional with weights in the unit of travel time, ranging from $1\sim 6$ sec. The numbers chosen here correspond to the time distances between nodes in realistic road networks. In all of our simulations, the rate of demand $\mathcal D\left(t\right)$ is constant with random OD pair generation. Given a specific demand, we can thus calculate the average travel time (the sum of the waiting time and the trip time) of the commuters and how it depends on the number of taxis running on the road. An example is plotted in Fig.(\ref{f1}).

There are two regions or phases of dependence on the taxi number as shown in Fig.(\ref{f1}). In the oversaturated region, the demand exceeds the supply of the taxis, and the average travel time decreases expotentially as the number of taxis increases. Note that here the average travel time can be very long due to long waiting time. In reality commuters will just quit looking for taxis if the waiting time is too long (thereby reducing the rate of demand in a time-dependent way), and that will mask some of the fundamental properties of taxi dynamics, which we aim to reveal here with a constant rate of demand. In the undersaturated region, there are excess taxis available to meet the transportation demand, and the average travel time decreases only sub-linearly with increasing taxi number. One should note that on average, the average trip time should not depend on the taxi number, because it is only determined by the average distance between the origin and destination of each OD pair. Increasing the taxi number only reduces the waiting time. In the undersaturated region, there are always empty taxis available when a potential commuter is generated. The sub-linear decrease of the average waiting time in the undersaturated phase is mainly due to the increase in the density of available taxis, which scales on the order of the square root of the taxi number for the finite system. Thus increasing the taxi number only reduces the average distance between the empty taxis and the location of the potential commuter, leading to a much smaller marginal reduction of the waiting time. 

We can thus define the boundary between the oversaturated and undersaturated region as the optimal number $N^*$ of the taxi, and $N^*$ depends on the factors including the demand, the underlying road network, as well as TRS which we will show later. When the taxi number $N<N^*$, adding more taxis can drastically improve commuter experience by reducing the travel time exponentially; on the other hand if $N>N^*$, the marginal benefits for the commuters are small when more taxis are added to the traffic system.

\begin{figure}[htb]
\centering
\includegraphics[width=.5\textwidth]{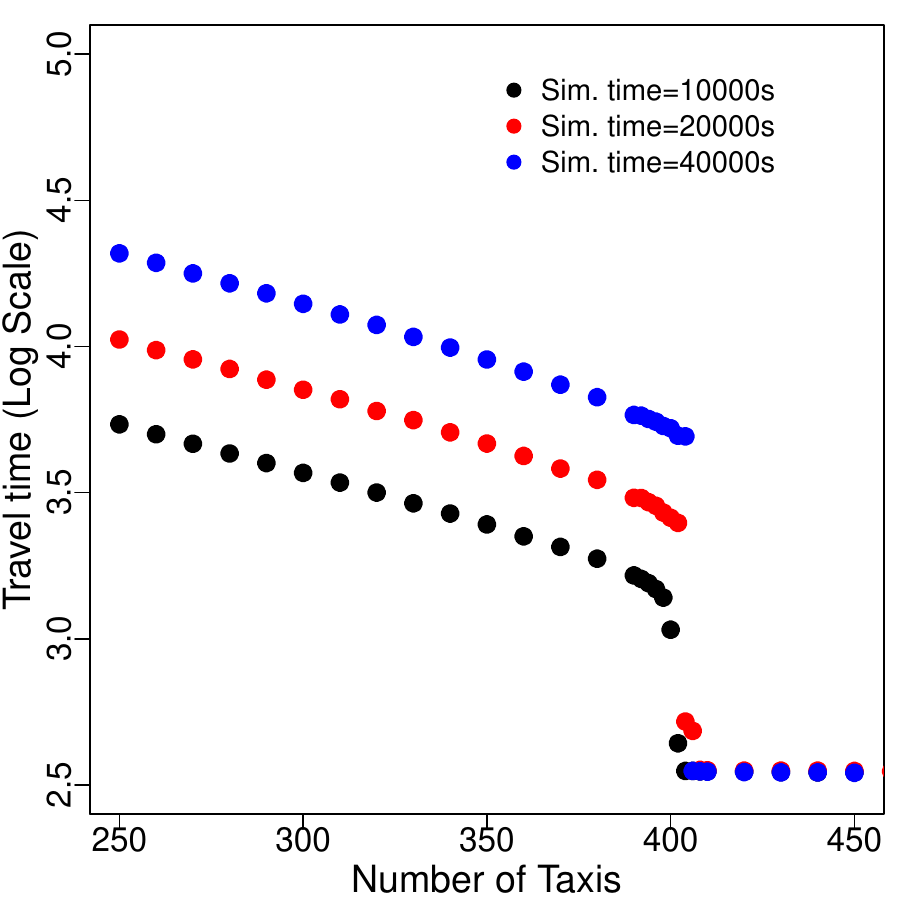}
\caption{Plots of the log of the average travel time against the number of taxis, with a fixed demand of one taxi request every second, with different simulation time. In this theoretical simulation we keep the commuters in the simulation no matter how long they will have to wait, to show that the phase boundary can be a well-defined number. The oversaturated phase is given by travel time depending on the simulation time, while the undersaturated phase is given by travel time not dependent on the simulation time (also see Fig.(\ref{heuristic}) for an intuitive understanding). For the more realistic case where commuters stop waiting after a certain cut-off time, the transition from one phase to another will not be sharp, and the same plots will be much smoother.}
\label{f2}
\end{figure}

Formally, $N^*$ can also be defined rigorously by varying the simulation time. The average travel time does not depend on the simulation time in the under-saturated region. In the over-saturated region, on the other hand, the longer simulation time leads to more commuters queueing for the taxi pickups, since all taxis are occupied almost all the time. This feature can be understood intuitively with the heuristic model described in Fig.(\ref{heuristic} a,b) in the oversaturated region. In Fig.(\ref{f2}), one can clearly see that $N^*$ is the unique number that separates two qualitatively different behaviours when simulation time is increased. Note that in Fig.(\ref{f2}) we have a much greater $N^*$ as compared to that in Fig.(\ref{f1}), because for the former the demand of the taxi is much greater.

\begin{figure}[htb]
\centering
\includegraphics[width=12cm]{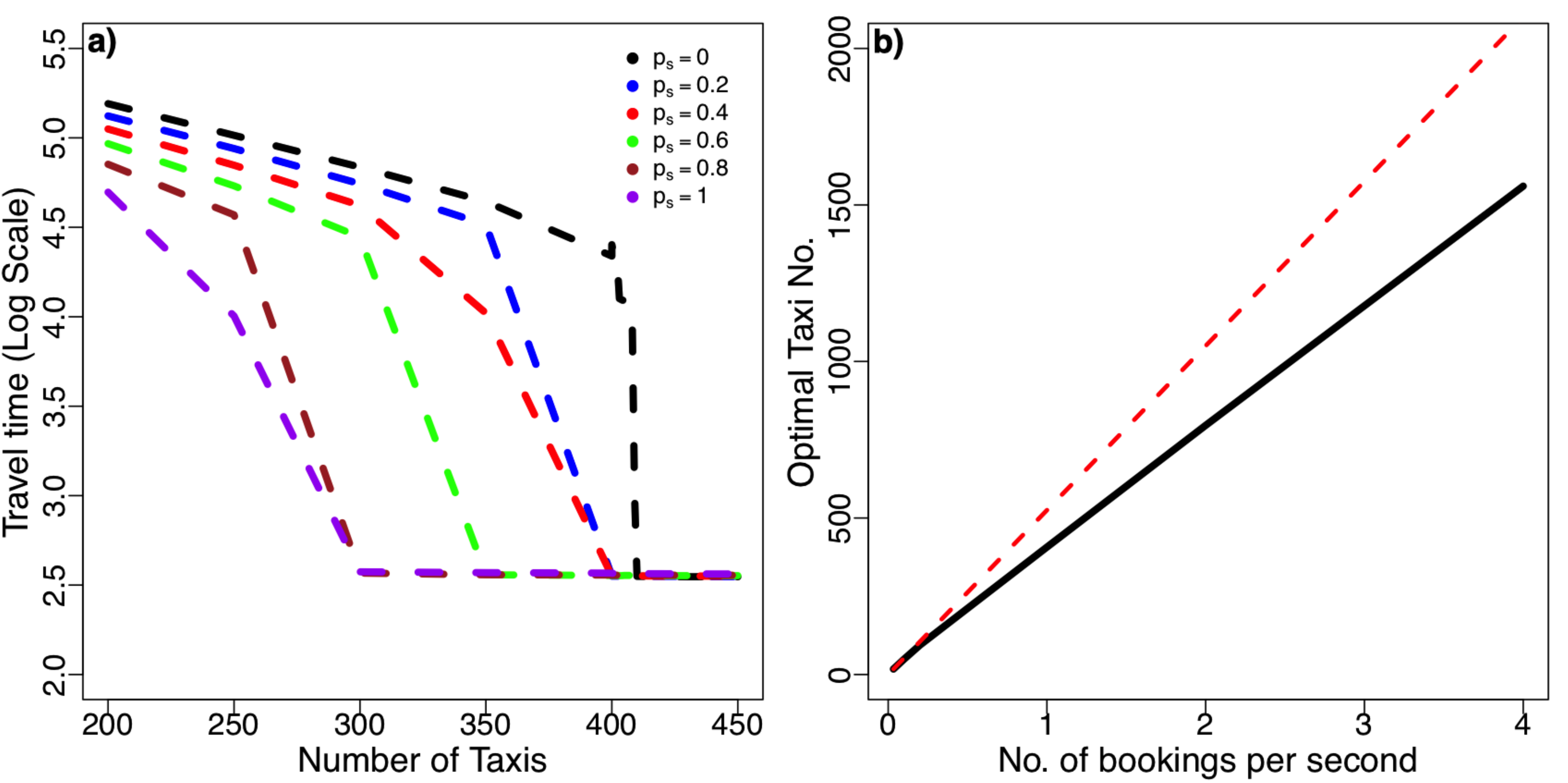}
\caption{a). The shift of the optimal taxi no. $N^*$ with different values of $p_s$. $N^*$ decreases monotonically but not in a smooth manner; b). The solid curve is the dependence of $N^*$ on the taxi demand, and the dotted curve is the straight line tangent to the solid curve near the origin, showing that the dependence of $N^*$ on the taxi demand is sub-linear. }
\label{f3}
\end{figure}

Intuitively, increasing the demand for taxi will always increase $N^*$, and in Fig.(\ref{f3}b) we can see that such relationship is linear when the demand is small, but sub-linear for large demand (probably due to the finite size of the road network). We now look at how $N^*$ depends on the transportation demand and the probability of taxi sharing (for each commuter generated in the simulation) by tuning $p_s$ from $0$ to $1$. We also find that $N^*$ decreases monotonically with increasing $p_s$. However, the rate of decrease is not linear and could be quite random for different demand patterns and different road networks (see Fig.(\ref{f3}a) as one example).  One should note that even if the commuter chooses to accept a shared trip (the corresponding taxi-sharing flag in $P_t^\alpha$ is TRUE), the actual trip may or may not be shared with another commuters, since it depends on if the nearest available taxi is occupied or not; if it is occupied taxi, it also depends on if the routes can satisfy the route matching algorithm.

\begin{figure}[htb]
\centering
\includegraphics[width=14cm]{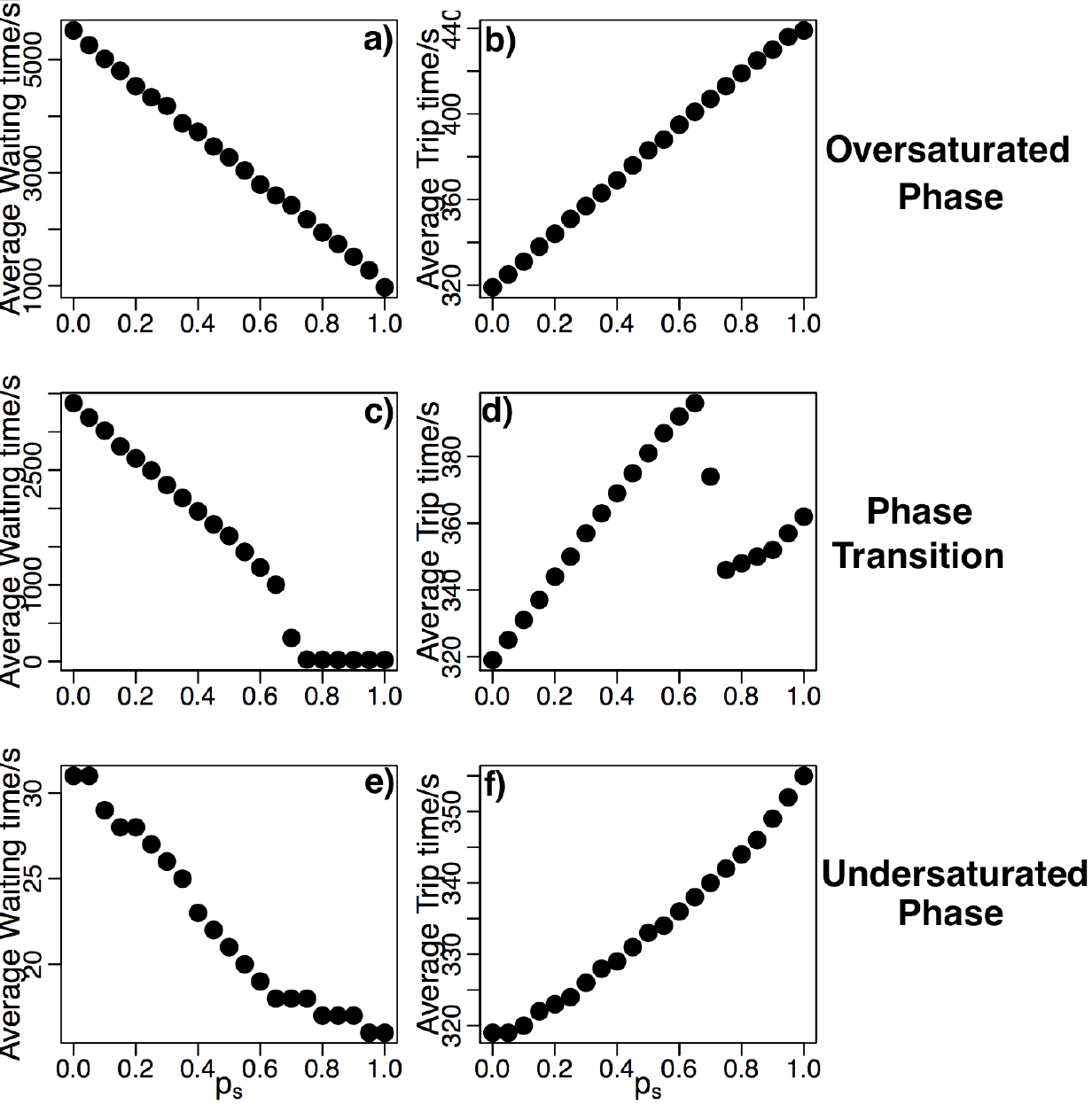}
\caption{a). Dependence of the average waiting time on commuter's willingness to taxi-sharing, in the over-saturated region; b). Dependence of the average trip time on commuter's willingness to taxi-sharing, in the over-saturated region; c). Dependence of the average waiting time on commuter's willingness to taxi-sharing, with the transition from over-saturation to undersaturation; d). Dependence of the average trip time on commuter's willingness to taxi-sharing, with the transition from over-saturation to undersaturation; e). Dependence of the average waiting time on commuter's willingness to taxi-sharing, in the under-saturated region; f). Dependence of the average trip time on commuter's willingness to taxi-sharing, in the over-saturated region.}
\label{sharedependence1}
\end{figure}

It is also useful to look at the dependence of the waiting time and trip time separately on the taxi number $N_T$ and commuters' willingness to share, $p_s$. For fixed demand and $p_s$, the average trip time is independent of the number of taxis, so the average waiting time and the average travel time depends on $N_T$ in qualitatively the same way.\cite{MATSim3,WaitingTime,WaitingTimeSupply} The dependence on $p_s$ is more interesting. In the over-saturated phase,  the average waiting time decreases rapidly with increasing $p_s$ (See Fig.(\ref{sharedependence1}a)). Phase transitions can also be observed at fixed supply and demand, when increasing $p_s$ pushes the taxi system from the over-saturated phase to the under-saturated phase, with a discontinuous drop of the waiting time (Fig.(\ref{sharedependence1}c)). This is because at small $p_s$ the number of taxis is not enough to cater to the taxi demand, but at large $p_s$ the same demand can be met with the same number of taxis with TRS. 

While average trip time increases with $p_s$ due to the increase amount of detours for pick-ups and drop-offs of the shared parties, there is a corresponding drop of the average trip time at the $p_s-$driven phase transition as well. This is because in the over-saturated phase, there are hardly any empty taxis around; in the under-saturated phase, on the other hand, the presence of empty taxis significantly reduces the actual number of shared trips, even when people are more willing to share (Fig.(\ref{sharedependence1}d)). With fewer shared trips, the average trip time is also lower due to fewer detours. When the traffic system is already under-saturated at $p_s=0$, the decrease of the average waiting time with respect to increase in $p_s$  is much less significant, especially when compared to the increase in the average trip time(Fig.(\ref{sharedependence1}e,f)). 

We can draw one important conclusion from the analysis of Fig.(\ref{sharedependence1}). It is normally considered that choosing to share the taxi trip would imply a longer travel time from one's origin to destination, which is the cost in exchange for cheaper taxi fare. Thus TRS is not recommended for commuters who are in a hurry. On the other hand, the total travel time also includes the waiting time for the taxi. When the supply and demand of the taxi is in the over-saturated region or close to the transition between the over-saturated and under-saturated region (i.e. the number of taxis running on the road is smaller or close to the optimal taxi number), choosing to share the taxi trip can potentially reduce the waiting time drastically. Even though the trip time will always be longer with TRS, the overall travel time can still be significantly reduced due to the reduction of the waiting time. For a city that maintains a total number of taxis close to the optimal number of taxis, TRS can thus be quite desirable from the commuters' perspective, especially during peak hours or bad weathers, as long as there are enough people willing to ride-share (i.e. $p_s$ is large enough to induce the phase transition).

\begin{figure}[htb]
\centering
\includegraphics[width=10cm]{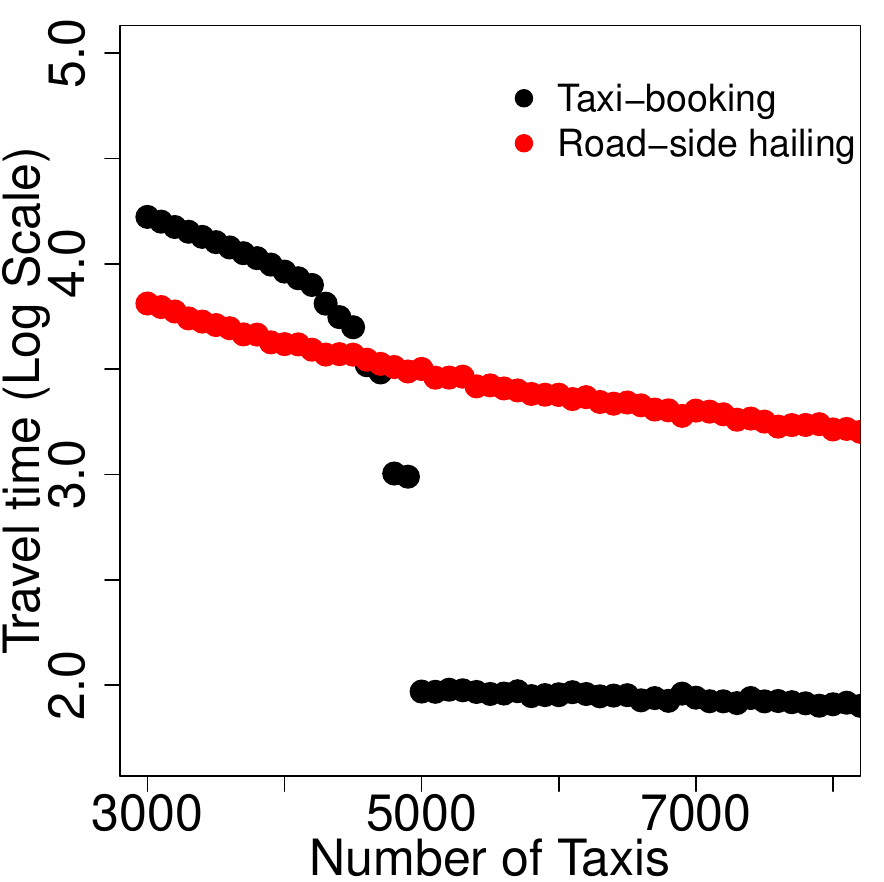}
\caption{Comparison of the taxi dynamics between a). Taxi-booking and b). Road-side hailing. The former has a clear kink and phae transition from the oversaturated region to the undersaturated region. For the latter, the dependence of average traveil time on the taxi number is a smooth curve that persists to very large number of taxis (not plotted here). In the simulation, the road-side hailing is implemented with random roaming of empty taxis.}
\label{comparison}
\end{figure}

Most of the interesting observations from this section comes from the phase separation of the taxi dynamics as shown in Fig.(\ref{f1}) and Fig.(\ref{f2}). To illustrate the non-trivialness of this aspects, we show here that such phase separation only exists with taxi-booking. If no taxi-booking is possible (i.e. back in the olden days when taxis can only be hailed from road-side), the total travel time is a smooth function of the taxi number, with no kinks or discontinuities even in the limit of large simulation time or large number of nodes in the road network; thus in this case $N^*$ is no longer well-defined. A comparison of different taxi dynamics can be seen in Fig.(\ref{comparison}). In particular, while in the undersaturated phase for taxi-booking, the average travel time does not increase with simulation time, for road-side hailing the average travel time will always increase with the simulation time for any taxi numbers (in the limit of large number of nodes). The random roaming of taxis implies that there are always an appreciable number of ``unlucky nodes" where all taxis happen to visit very rarely, since unlike taxi-booking, there is no exact information on where the waiting commuters are. From the heuristic point of view of Sec.~\ref{heuristicmodel}, the ``outflow" of the funnel increases linearly with the number of taxis with taxi-booking, while with road-side hailing it only approaches asymptotically to the ``inflow" in the limit of large taxi numbers. Thus taxi dynamics with taxi-booking tends to be much more interesting, particularly with our focus on TRS, where taxi-booking is a prerequisite. 

\subsection{The Road Network of Singapore}

\begin{figure}[htb]
\centering
\includegraphics[width=12cm]{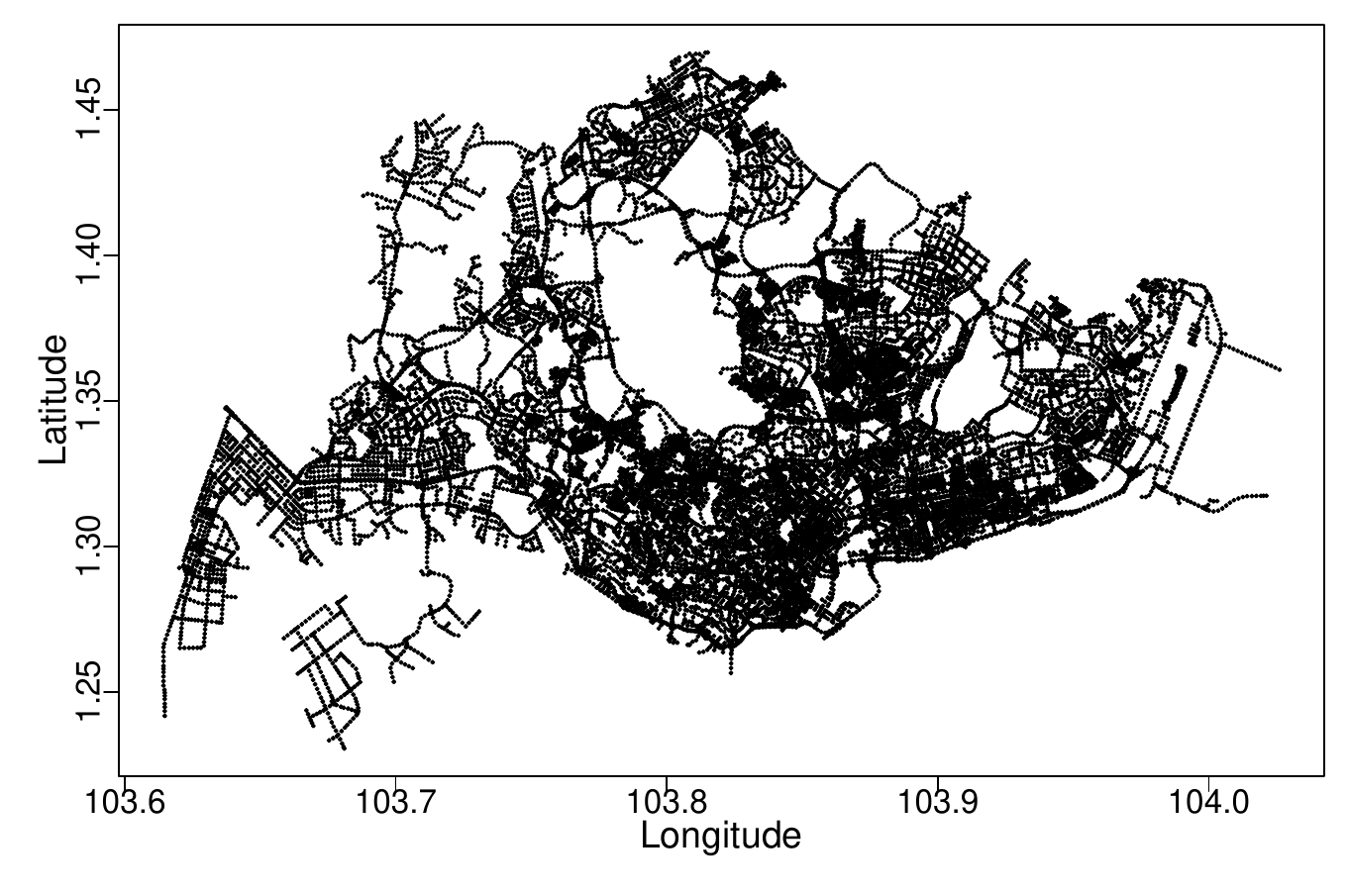}
\caption{Geographic distribution of the road network of Singapore, which contains 27185 nodes. Each dot in the graph is a node. The two nodes are connected by an edge of it is possible for the vehicle to travel from one of the node to the other. The weight of the edge is the time it takes for the vehicle to travel from one node to another. It could be static and calculated from the geographic distance divided by the road speed limit, or it can be dynamical based on the real-time traffic condition.}
\label{f5}
\end{figure}

We now look at the realistic road network of Singapore, extracted from the open street map with nodes as locations along the road accessible to motor vehicles and spaced tens of meters apart. Geographically, the distribution of the nodes are shown in Fig.(\ref{f5}), though for numerical simulation the geographical locations of the nodes are not important, and all the relevant information is stored in the nodes and edge weights of the network. We again generate the demand of the taxis randomly, with the probability of the origin and destination of each trip evenly distributed over all the nodes.

\begin{figure}[htb]
\centering
\includegraphics[width=12cm]{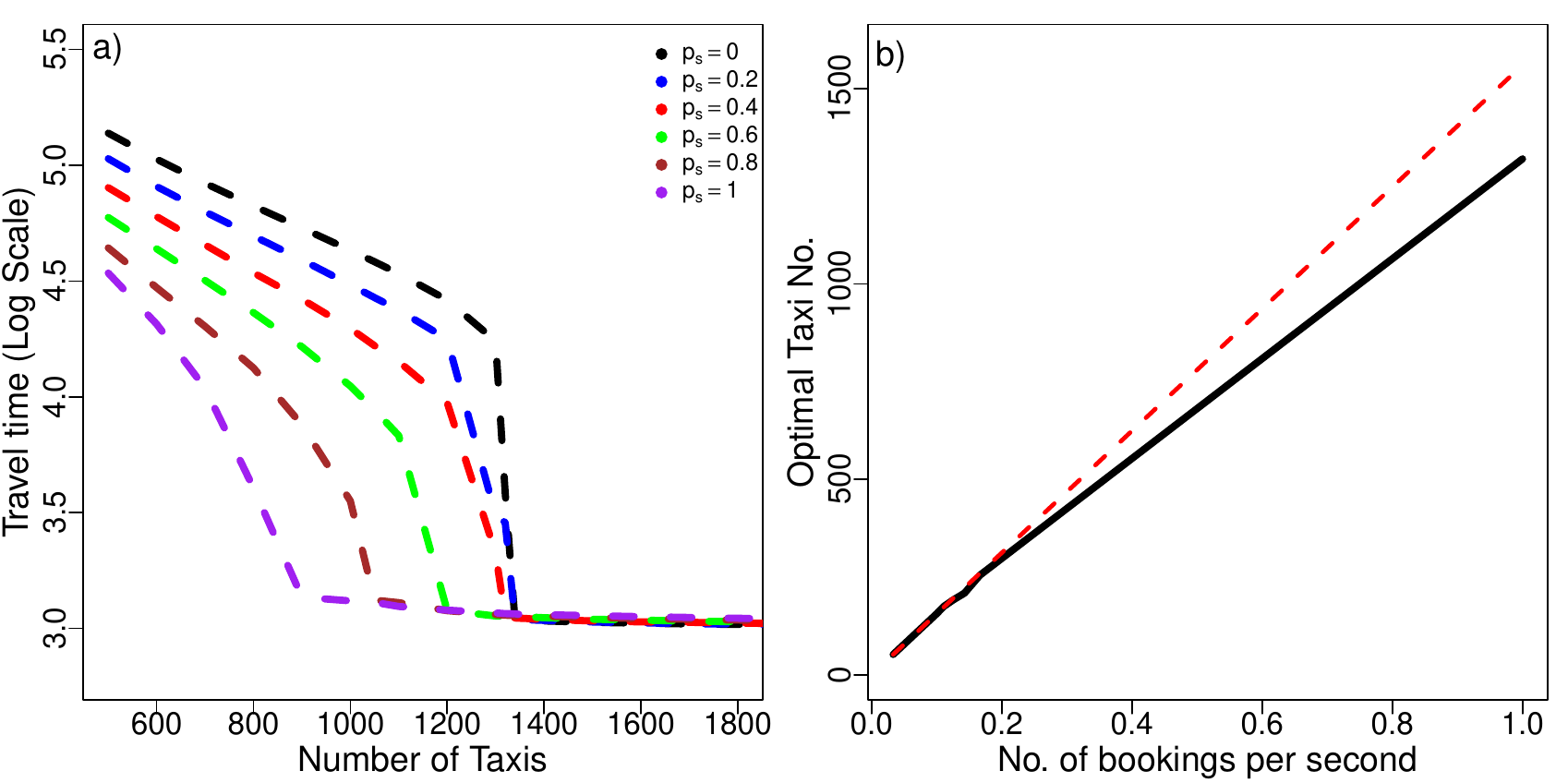}
\caption{Dependence of $N^*$ when we vary $p_s$ or the taxi demand, similar to that of Fig.(\ref{f3}) but with the road network of Singapore.}
\label{f6}
\end{figure}

Most of the analysis and qualitative features for the artificial network in the previous section can also be applied for realistic networks like this one. The optimal number of taxis can be defined identically, and its dependences on the demand and willing to share are qualitatively the same, which are plotted in Fig.(\ref{f6}).

\begin{figure}[htb]
\centering
\includegraphics[width=14cm]{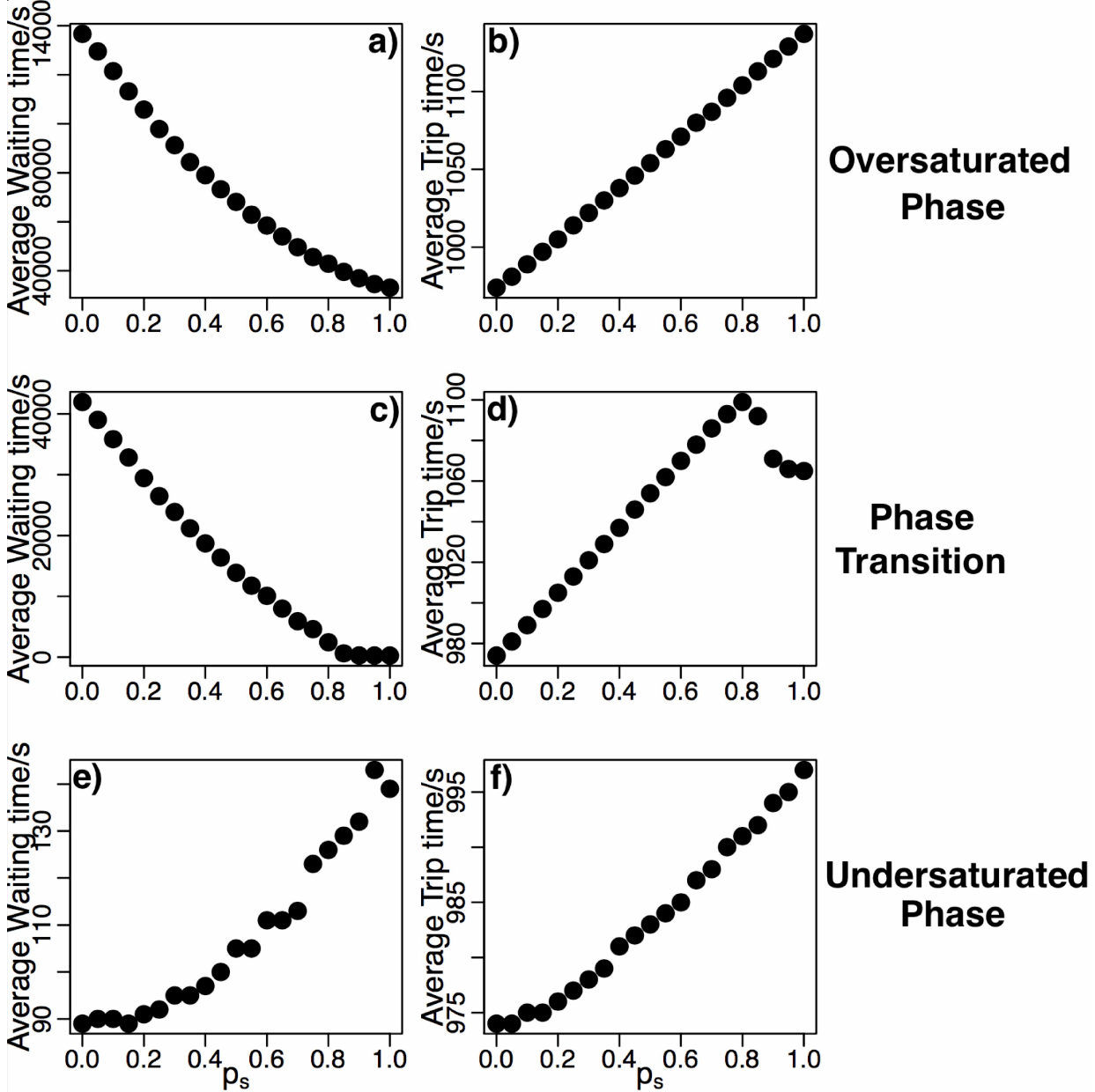}
\caption{Dependence of the average waiting time and trip time on commuter's willingness to share, both in undersaturated and oversaturated regions similar to Fig.(\ref{sharedependence1}) but with the road network of Singapore.}
\label{sharedependence2}
\end{figure}

The dependence of the average waiting time and average trip time on the willingness to taxi-sharing are also plotted in Fig.(\ref{sharedependence2}). While most of the features are qualitatively the same as Fig.(\ref{sharedependence1}) from the artificial network, one notable difference is particularly interesting and illustrating how the topology of the underlying road network affects the taxi dynamics. In Fig.(\ref{sharedependence2}e), the increase of the average waiting time on the willingness to share in the under-saturated region is counter-intuitive. Given that the pick-up and drop-off points of the taxi trips are quite random, one would expect that for each commuter booking for a taxi at a certain location, allowing taxi-sharing always increase the number of potential taxis coming to pick him/her up. Since the taxi closest to the commuter will come for the pick-up based on our simulation algorithm, on average one would always expect a shorter, if not the same, waiting time for commuters choosing to accept taxi-sharing. While this heuristic argument agrees with Fig.(\ref{sharedependence1}e), it does not agree with Fig.(\ref{sharedependence2}e).

One possible explanation could be that TRS alters the taxi dynamics in terms of the statistical spatiotemporal distribution of the occupied and empty taxis: even though the origins and destinations of the trips are random, TRS makes taxis available for pick-up further away from randomly generated booking locations (which are also close to the origins of the taxi trips) in a subtle way. Since all the nodes are equivalent in our stochastic simulation, the distribution of the edges connecting the nodes should be responsible for the paradoxical behaviour that does not occur in the artificial network.
\begin{figure}[htb]
\centering
\includegraphics[width=12cm]{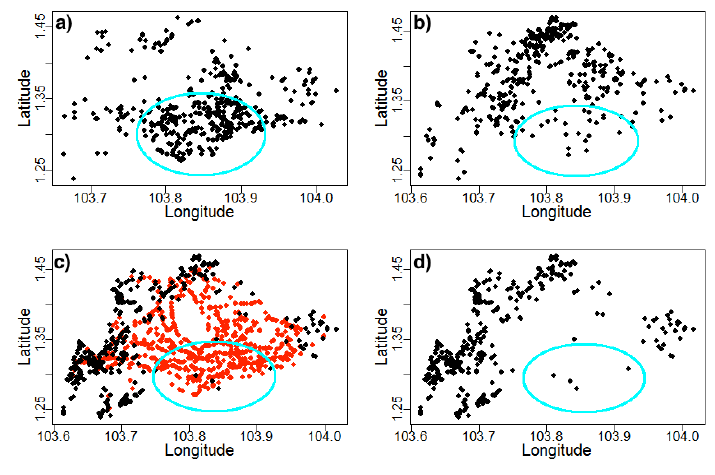}
\caption{a). Numerical simulation with no taxi-sharing, with the locations of the empty taxis at a time snapshot early in the simulation; b) Numerical simulation with no taxi-sharing, with the locations of the empty taxis at a time snapshot at simulation time more than $10^6$ sec; c).Numerical simulation where commuters' taxi-sharing willingness is 1 (as long as there is a good route-matching between two commuters, their trips will be shared), with the locations of the empty taxis at a time snapshot after the simulation time of $10^5$ sec. The black dots are empty taxis, the grey (or red with color online) dots are occupied taxis with one commuter (they are available for another commuter as long as the booking algorithm and the route matching algorithm permit; d). same as c) but plotting only the empty taxis. The circle indicates the location of the city center (CBD area).}
\label{f8}
\end{figure}

As we can see from Fig.(\ref{f8}), if we do not have TRS, there are always a large number of empty taxis in and around the city center, since we are in the under-saturated phase. When most people are open to TRS, the number of available taxis increases significantly, but in and around the city center, most of the available taxis are actually occupied taxis with commuters who are open to TRS; there is a conspicuous lack of \emph{empty} taxis in and around the city center. Given the high density of nodes in and around the city center, the average waiting time is strongly affected by people booking from this region. This is because even though there are many available (yet already with one passenger) taxis, the chances that the routes of these taxis can be matched to the route of the new booking are always quite low, so most of the occupied taxis open to TRS are effectively unavailable. On the other hand, empty taxis do not have this constraint of route matching. Thus empirically, the waiting time for the commuters in and around the city center could be longer, since the effective available taxis, after the filtering of route matching, is actually lower than the case when no one is allowed to do taxi-sharing.

It is thus quite interesting that with the road network of Singapore, taxi-sharing tends to push a lot of the empty taxis to the city fringe (including the industrial areas). Note this is independent of the spatial variation of the booking demand patterns, since each node is treated equally when random generation of the OD pairs are concerned; it is thus a purely network effect. Since in our simulations empty taxis roam randomly, when taxis drop off commuters at the city fringe and becomes empty, they tend to stay around there because of the lack of nearby demands as a result of the low density of nodes (see Fig.(\ref{f5})). On the other hands, vehicles in and around the city centers get less chance of being empty, because taxi-sharing and route matching lead to vehicles moving around more with commuters on board. While in reality such effects could be suppressed by a non-trivial $\mathcal R_e$ (since after dropping off commuters near city fringe the empty taxis naturally tend to move towards the city center) thus probably \emph{unobservable}, what the simulation reveals here is one aspect of potential ``penalty" for taxis carrying commuters towards less popular areas especially from the perspective of TRS, and could be suggestive on how to optimise the routing strategies for the drivers, especially when TRS becomes prevalent.

\section{Simulations with Empirical Origin-Destination Data}\label{empiricalanalysis}

In the previous section, we analysed some of the qualitative features of the taxi dynamics based on simple models with no empirical input. For more quantitative simulations and predictions on the detailed interactions between commuters and taxis, we need empirical input on the actual number of taxis on the street, the empirical spatiotemporal distribution of the origin-destination (OD) pairs, actual route choices (which may differ from the shortest paths between the origin and destinations), as well as other important details including road-side hailing, dynamics of empty taxis, taxi stands and the actual booking assignments, etc. A detailed study with such empirical data, especially within the city of Singapore, is work in progress and we will report the results elsewhere.

In this section, we show the results of simulations with the empirical road network of Singapore, and partial empirical input consisting of the actual demand of taxis, in terms of realistic spatiotemporal distribution of the OD pairs. Most of the results are qualitatively the same as discussed in the previous section where we employed an artificial or Singapore road network and artificial taxi demand (with a constant rate of random OD pair generation); this illustrates the robustness of our models that are capable of capturing the essential universal features of taxi dynamics. We highlight in Fig.(\ref{empiricalot}) that with more acceptance of TRS, as modelled by an increase in $p_s$, the optimal number of taxis needed on the street, given the empirical demand, can be reduced drastically.

\begin{figure}[htb]
\centering
\includegraphics[width=15cm]{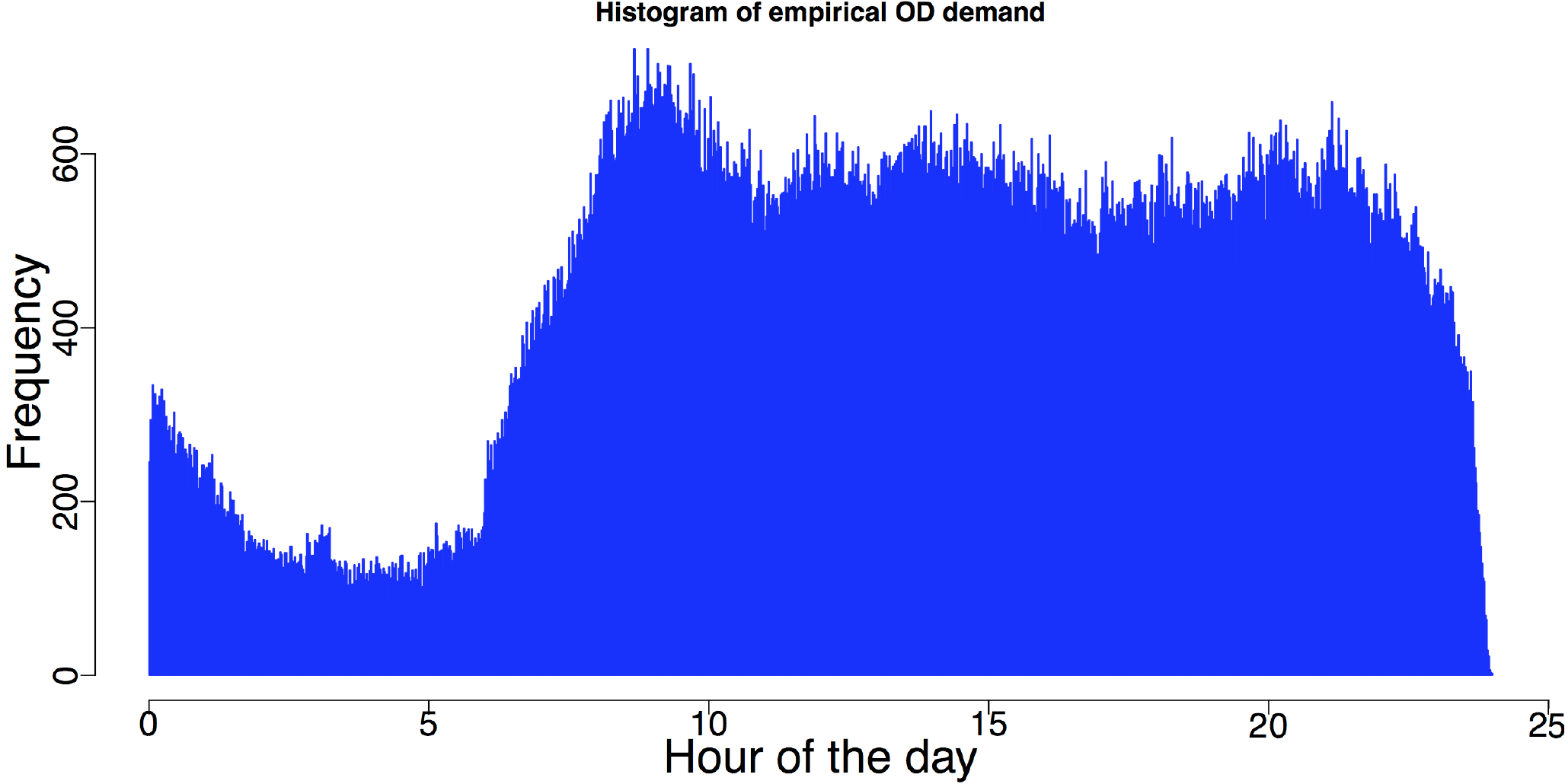}
\caption{The histogram of the empirical OD demand for a typical day in Singapore, with a total of 589243 taxi trips in 24 hours.}
\label{hist}
\end{figure}

We now discuss some of the quantitative differences between simulations with artificial and empirical input, which serves as the precursor for more practical studies in the future. On average, the empirical data shows that there are around seven trips or OD pairs per second, with temporal distribution shown in Fig.(\ref{hist}). The average distance between the origin and destination from all empirical OD pairs is around 9 km.

\begin{figure}[htb]
\centering
\includegraphics[width=15cm]{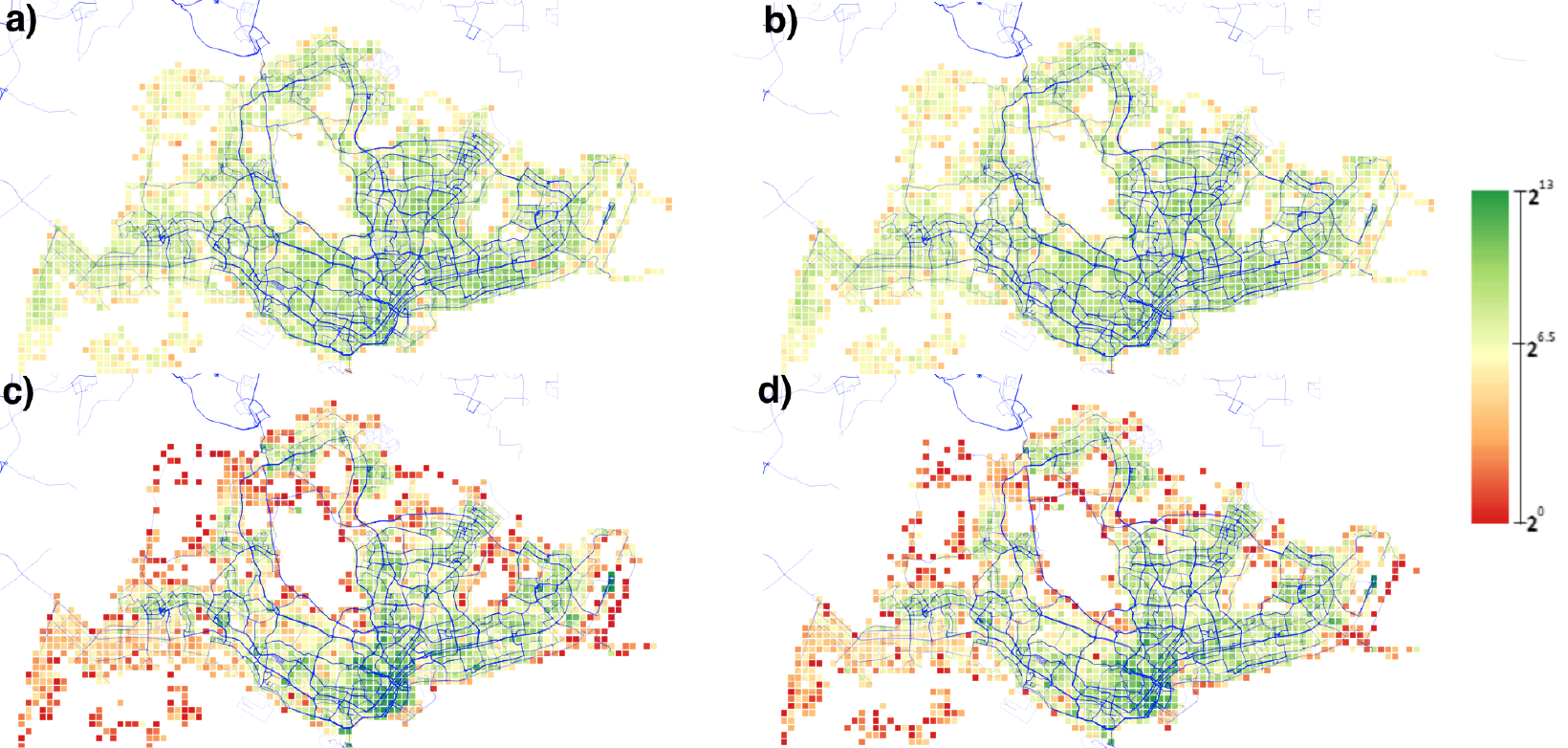}
\caption{Geographic distribution of a) origin and b) demand over one day, with random generation of the OD pairs; Geographic distribution of c) origin and d) demand of the taxis over one day of the empirical data. The color scale gives the number of origin/destination sites in each square.}
\label{empirical}
\end{figure}

\begin{figure}[htb]
\includegraphics[width=12cm]{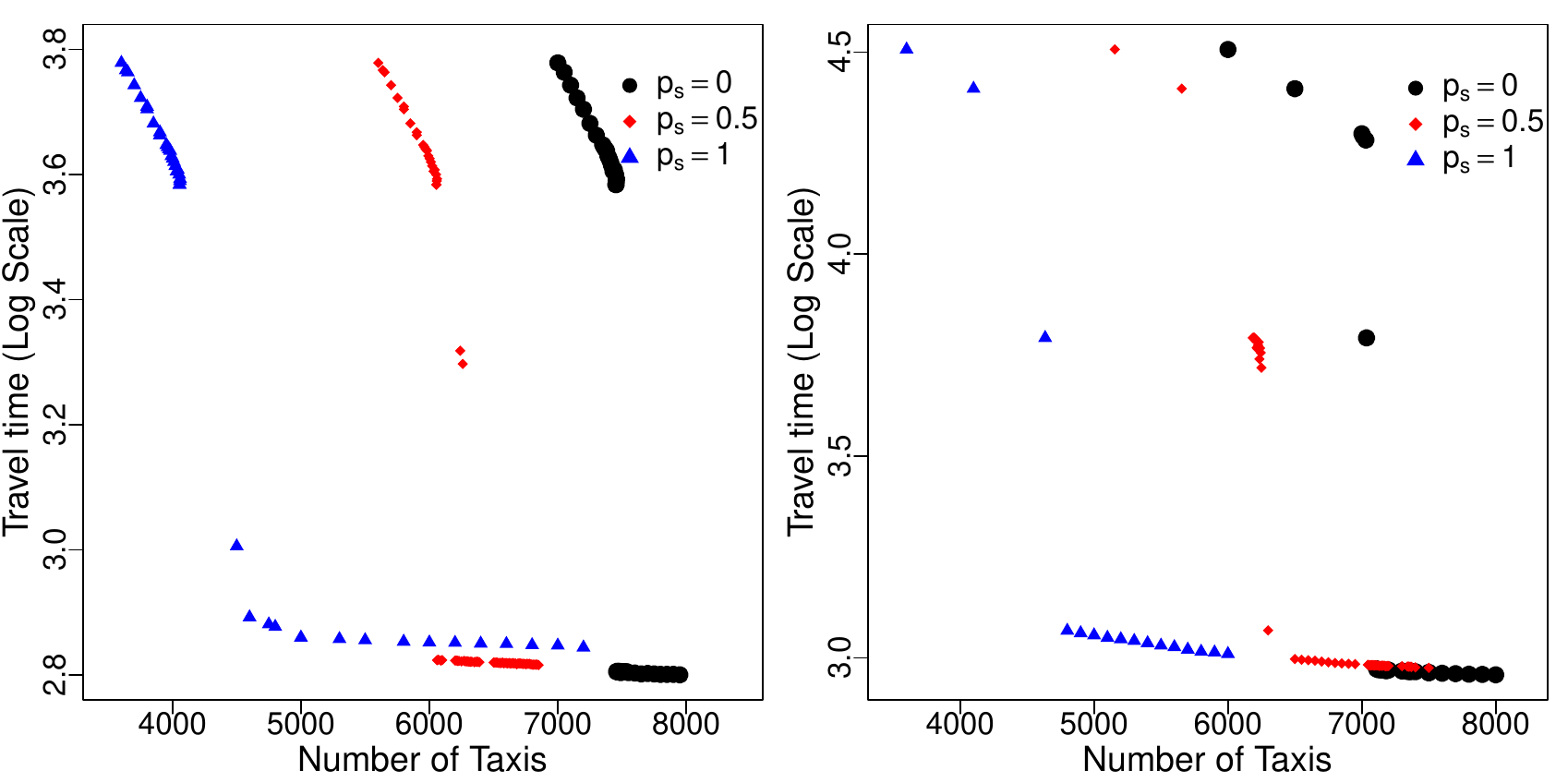}
\caption{Travel time plot at different $p_s$, comparing the dynamics with empirical OD pair (left) and randomly generated OD pair (right) with the same average rate of $P_t^\alpha$ generation.}
\label{empiricalot}
\end{figure}

It is interesting to compare the dynamics of the empirical data with a randomly generated OD pairs over 24 hours, assuming that the OD pairs are generated \emph{uniformly} over time, with a constant rate of generation of around seven ODs per second. The simulation is done with the same road network over the same period of time, with the same total number of trips. The only differences are the temporal distribution of the OD pairs (non-trivial for empirical ODs based on Fig.(\ref{hist}), and uniform for the random data), as well as the spatial geographic distribution of the origins and destinations. For randomly generated OD pairs, each node has the equal chance of being either the origin or destination, under the constraint that the trip time between the origin and destination has to be at least five minutes. The comparison of the spatial distribution of the OD pairs are shown in Fig.(\ref{empirical}).

One can see that for random OD pairs, the locations are evenly distributed to all nodes, including those in the outskirts of the city. On the other hand, empirical data shows that few trips start or end in the outskirts, and most of the trip activities are concentrated in the city center. This clustering of origins and destinations is reflected by the average travel time per trip, which is around $550$ seconds with empirical OD pairs, as compared to around $900$ seconds from the random OD pairs. Intuitively, we would expect fewer taxis needed with the empirical data, because shorter travel time implies faster rate of ``outflow" (see Fig.(\ref{heuristic})), or a faster rate of empty taxis recovering from occupied ones. In contrast, numerical simulations show otherwise. As we can see in Fig.(\ref{empiricalot}), for $p_s=0$, the optimal number of taxis is actually greater for empirical OD pairs, as compared to the randomly generated one. This could only be explained by the fact that the random OD pairs have a different spatial and temporal distribution of the commuter generation. In particular, the clustering of the OD pairs geographically tends to make the traffic dynamics less efficient. 

Even though clustering of ODs imply more taxis are needed with the same demand, it is also shown clearly in Fig.(\ref{empiricalot}) that taxi-sharing has a greater effect with empirical ODs. This is intuitive, since route matching could be more efficient when trips are concentrated around CBD and residential areas according to the empirical data, as compared to more evenly distributed trips from random generation. Thus a higher adoption rate of taxi-sharing can more than compensate for the inefficiencies due to clustering of ODs, leading to shorter travel time and fewer number of taxis needed on the road. Such interesting interplay between the spatiotemporal distribution of the OD pairs and the underlying road network topology could be important from the perspective of urban planning.

\section{Application of Phase Boundary in Taxi System Optimization}\label{methodology}

The simulation platform we have constructed, together with the conceptual understandings and qualitative characterisations from the detailed analytical and numerical computations in this paper, can lead to development of various novel methodologies and applications in improving the efficiency of the complex taxi systems. In this section, we give a succinct discussion on the potential importance of the phase boundary, given by the optimal number of taxis $N^*$, used as a benchmark for the optimisation of the taxi systems.  A more detailed investigation devoted to taxi system optimisations with taxi-ridesharing will be published elsewhere given the limited scope of this work.

In principle, the optimal number of taxis $N^*$ can be detected empirically by large scale collections of the empirical data over an extended period of time, with precise control of the availability of the total number of taxis. When there are fewer taxis than a certain critical number, the average waiting time of commuters would increase drastically, indicating the onset that the supply of taxis falls below the demand. In practice this could be rather difficult to collect such data over a large scale, and a better option is to use agent based simulation platform (similar to the one we presented in this paper), combined with detailed and accurate empirical input (including the demand and taxi behaviours such as loading/unloading time for passengers), to estimate $N^*$ from numerical computations. This in generally can be quite viable even for quantitative estimates, given the rather sharp transition with taxi-booking we have seen from the simulations in previous sections.

Give a particular taxi system, the average waiting time $\bar w_T^*$, when there are $N^*$ taxis in the system, is also a well-defined quantity. The rather different dynamics between the oversaturated phase and the undersaturated phase as we have seen in Sec.~\ref{simulatedanalysis} and Sec.~\ref{empiricalanalysis} shows that both $N^*$ and $\bar w_T^*$ are the important benchmarks for system optimisation. One should note that $\bar w_T^*$ is intrinsic to the taxi system, given a specific demand pattern for the taxis, the infrastructure (i.e. the road network) and the taxi behaviours (e.g. routing). Conventionally, external targets or metrics are usually imposed on the taxi system to determine if the system is efficient enough. For example, we would like the average waiting time for commuters to be below $\bar w_t=5$ minutes, which is a reasonable target. It is, however, not obvious what is the best strategy in achieving that target. In particular, just increasing the number of taxis in the system could be the wrong choice if $\bar w_t<\bar w_T^*$. This is because in the undersaturated region, the marginal reduction of the average waiting time by increasing taxi numbers is small, so potentially a large number taxis need to be added to bring down the average waiting time. This could lead to undesirable consequences such as traffic congestions, unnecessary increase in fuel costs and green house gas emissions. In such cases, one should focus on other factors such as promoting ride-sharing, more intelligent routing of taxis, or even more optimal road network infrastructures (e.g. dedicated taxi stands) to shift $N^*$ and thus $\bar w_T^*$. Only in the oversaturated region when $\bar w_T^*<\bar w_t$, is tuning the total number of taxis the most effective way of optimising the taxi systems.
\begin{figure}[htb]
\centering
\includegraphics[width=12cm]{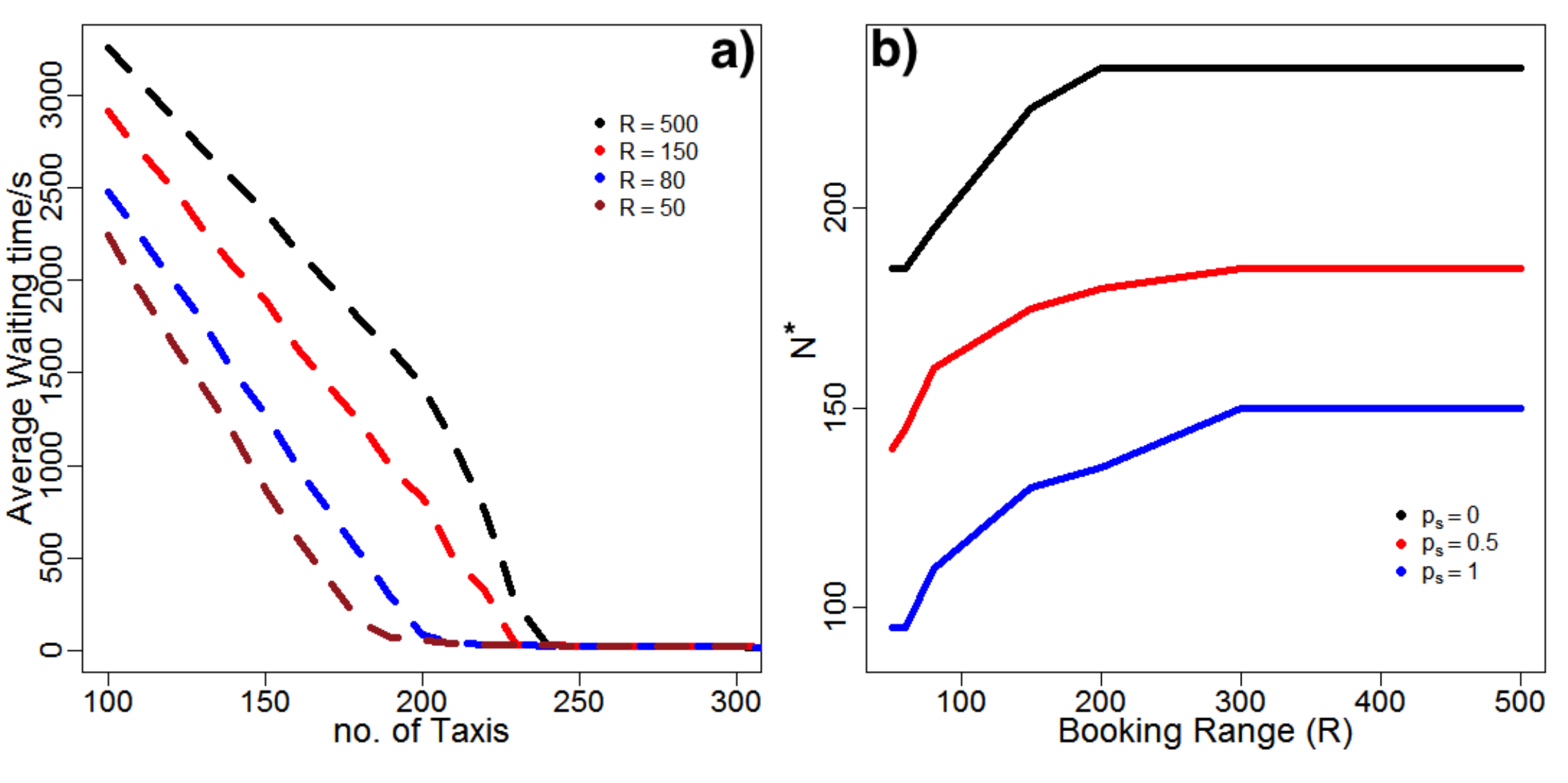}
\caption{a). Shifting of the phase boundary $N^*$ when the booking range $R$ is tuned, for the case where $p_s=0$; b). Different shifting behaviours of $N^*$ at different $p_s$.}
\label{fig17}
\end{figure} 

Examples of how $N^*$ or $\bar w_T^*$ depend on TRS or the demand patterns (which can also be implicitly dependent on road network and infrastructures) are shown in Fig.(\ref{f6}) and Fig.(\ref{empiricalot}). Obviously they also depend on the road network directly, as we can see from the contrast between the artificial road network and the empirical Singapore road network in Sec.~\ref{simulatedanalysis}. We also gave examples on how we can optimise $N^*$ and $\bar w_T^*$ via tuning certain aspects of taxi behaviours in Sec.~\ref{bookingrouting}. Here, we give another example to show that $N^*$ can be tuned by optimising the pick-up algorithm by changing the booking range $R$, as one can see in Fig.(\ref{fig17}). One should also note that tuning $R$ does not alter the qualitative behaviours of the taxi dynamics.

Optimising the pick-up algorithm generally involves not having to assign one of the available taxis immediately after the commuter sends the booking. Instead, one can have a buffer window within which several commuters send the booking requests. There can thus be an efficient assignment for all available taxis during this buffer window (including occupied taxis that are about to become empty after reaching destinations nearby), to each of the commuters sending requests within the same window, so as to minimise the overall waiting time of all commuters. Effectively, different algorithms lead to different average pick up range $R$. As far as the system dynamics is concerned, they are generally equivalent to the tuning of $R$ in our algorithm, as long as we scan for the closest available taxi with high enough frequency (every one second in our simulations). This is made possible with our highly efficient routing algorithm (see Sec.~\ref{routing}), allowing us to use $N^*$ as the benchmark to find the optimal range of booking (see Fig.(\ref{fig17})).

\section{Limitations and Universality of Numerical Simulations}\label{limits}

All models and simulations are approximations of the real world, and this is especially true for the taxi dynamics, embedded in a highly complex urban system. It is thus important to discuss in details the limits of our simulation analysis, as well as its universalities. For the former, we need to understand that certain detailed behaviours of the taxi dynamics cannot be captured by our simulations, and their quantitative (and even qualitative) results may not be directly compared to the empirical data. For the latter, we need to justify that many of the qualitative results from the numerical analysis are universal and can be generally applied to the empirical data of various urban systems, in spite of various simplifications we employed in our simulations. 

For commuter pick-up's, it is important to note that our simulation follows a booking algorithm, in which taxis near the pick-up points know the exact position of the pick-up, and the nearest available taxi will always accept the booking and start moving to the pick-up point right away. We thus do not focus on accommodating taxi-hailing on the street, and we do not accurately model the time it takes for each commuter to board the taxi before heading towards the destination. The drivers also do not have the choice of rejecting the assigned commuter. While other realistic details may affect the microscopic experience of individual taxis and commuters, the qualitative effects of supply and demand of the taxis on average waiting or trip times, etc. should not be affected, which we tested with extensive (though not necessarily thorough) simulations of various different microscopic details. When comparing to the empirical data, one should note that the rate of commuter generation and the number of taxis in the simulation are both parameters indicating the effective demand and supply, and should not be interpreted literarily. For example, the actual optimal number of taxis in a city would deviate from the $N^*$ obtained from the simulation, since occasionally taxi drivers may refuse to accept the assigned booking or be unable to find the commuter (which tends to increase $N^*$), while road-side taxi-hailing may increase the efficiency of the pick-up's (which tends to lower $N^*$). Nevertheless, the characteristic dependence of the taxi dynamics on the increase/decrease of taxi supply/demand based on simulation should still be useful for predicting similar characteristics from the empirical data.

The movement of the empty taxis after they drop off the last commuter is an important part of the taxi dynamics, and in reality different taxi drivers may have different strategies (mostly based on experience) in maximizing the possibility of picking up the next commuter as early as possible. Such strategies can depend on different time of the day as well as different locations within the city. In our simulation, this part of the taxi dynamics is simplified. Empty taxis will just roam randomly from the destination of the last trip. Assuming in reality the drivers are on average more experienced, the simplification in the simulation implies the simulated average waiting time could be longer, and the simulation could over-estimate the waiting time and $N^*$. On the other hand, we do not expect the actual routing strategy of the empty taxis to alter the qualitative behaviours of most of the simulated results, even if the strategies are vastly superior than a random routing.

Many aspects of the dynamics of taxi-sharing strongly depend on the actual booking and route-matching algorithm, that determine whether or not two trips will be shared. In our simulation, a simple greedy algorithm is used, so that two routes will be shared if the detour time for each of the shared party will not exceed a certain cut-off time based on the shortest path routing, and the cut-off time is the only parameter in our algorithm. In addition to that, we also introduce the parameter of the commuters' willingness to share their ride, which is the probability that the trip will be shared, given that all other criteria for route-matching based on the stated algorithm are satisfied. We believe such a simple model is sufficient in capturing the most salient effects of taxi-sharing on taxi dynamics, and the effects could be universal. This is because the taxi dynamics is mainly affected by the percentage of trips that actually shared, especially if we ignore detailed individual-level experience of taxi-sharing. Different algorithms and different parameters in the algorithm will just alter the percentage of shared trips, given the same level of willingness for the commuters to share. Plots in Fig.(\ref{sharedependence1}) and Fig.(\ref{sharedependence2}) may thus be shifted along the x-axis (for example if we allow the detour time to be longer for longer trips), but the overall dependence and trend of the plots should be qualitatively the same. 

We also simplify our simulation by limiting each $P_t^\alpha$ to contain only one commuter, and each taxi to have a maximum of two commuters on board. Such simplification can be easily relaxed, and even generalized to the cases for adaptive shuttles with vehicles in the form of mini-van or buses. If three or more $P_t^\alpha$ can share the ride, the routing matching algorithm and the simulation will be technically more complicated but conceptually equivalent to the cases we simulated in this paper. While detailed studies of real-time adaptive shuttle services based on our simulation platform will be presented elsewhere, for taxi services our extensive numerical simulation shows that this simplification does not affect the universal features discussed in this paper.

\section{Summary and Outlooks}\label{conclusions}

In summary, we have presented the mathematical formulation of the different components of the taxi system that are integrated into our simulation platform, and performed extensive numerical simulations to analyse how the interplay between the supply and demand of the taxis, the complexity of the road networks, the spatiotemporal distribution of the taxi demand, as well as the acceptance of taxi-sharing, can affect both the commuters' and drivers' experience as a result of the characteristic taxi dynamics. While we employ a minimalist approach in most of the simulations performed in this paper, our simulation is versatile and can incorporate much detailed behaviours from the empirical data or from sophisticated models on commuter generation, taxi routing and taxi-sharing. It can also be readily extended to study and evaluate the performance of adaptive shuttle services, in which vehicle occupancy can be large and ride-sharing can involve multiple parties.

We discussed several key universal features of the taxi system that are captured with our minimalist models and present in more realistic and sophisticated simulations. There are two phases of the taxi dynamics: in the oversaturated phase when there are more demand for the taxi services than the supply of taxis, changing the number of taxis on the road will affect the average commuter waiting time exponentially; in the undersaturated phase when there are more taxis on the road than the demand for taxi services, changing the taxi numbers only affect the average commuter waiting time sub-linearly. The boundary between the oversaturated and undersaturated phase allows us to meaningfully define an optimal number of taxis given an urban setting, including the underlying road network and the way commuters interact with taxis.

We also show that this optimal number of taxis depend on the details of the road network as well as how the origins and destinations of the taxi trips distribute spatially and temporally. With more accurate empirical data we can thus offer good quantitative guidance on how to regulate the size of the taxi fleet in the city. In particular, we show that taxi-sharing can significantly reduce the optimal number of taxis. In addition, we show that in the undersaturated phase, choosing to share the taxi ride leads to longer travel time, since there is an increase of trip time due to detour, but very small decrease in waiting time; in contrast, in the oversaturated phase, which could be induced by the lack of taxis, surge of demand due to bad weather, peak hours or public transportation disruptions, choosing to share the taxi ride can drastically reduce the waiting time. Thus, even though the trip time itself could be longer, the overall travel time (which is the sum of the waiting and trip time) can be significantly shorter. By maintaining an optimal taxi fleet size and given the empirical fluctuation of the demand, we can thus show that taxi-sharing can be both a cost-saving and time-saving behaviour for the commuters.

It is possible for our simulation platform to perform further detailed and quantitative analysis from the drivers' perspective, especially on developing good routing algorithms both for occupied and empty taxis, and using the platform to benchmark  and evaluate such algorithms. With real-time traffic data on the road network, we can also undertake real-time simulations with a time-dependent network in which the weight of every edge depends on the real-time traffic condition and thus the average velocity of vehicles along that edge. Built on our current capabilities, we will also look to integrate the taxi system with the public transportations including buses and trains, for the goal of developing optimal management strategy for bridging services to complement public transportations especially during peak hours, and to make buses and trains more accessible via the first and last mile services.

\section{Acknowledgements}
This research was partially supported by Singapore A$^{\star}$STAR SERC ``Complex Systems" Research Programme grant 1224504056. We also thank the referees for their very insightful comments during the review processes.

\end{document}